\documentclass[12pt]{article}

\usepackage{latexsym}
\usepackage{amsfonts}
\usepackage{amssymb}
\usepackage{amsmath}

\usepackage{cite}

\font\tenmsbm=msbm10 scaled 1200
\font\sevenmsbm=msbm9
\newfam\msbmfam
\textfont\msbmfam=\tenmsbm \scriptfont\msbmfam=\sevenmsbm

\makeatletter
\@addtoreset{equation}{section}
\makeatother

\newcommand{\eref}[1]{(\ref{#1})}

\def\be{\begin{equation}}
\def\ee{\end{equation}}
\def\ba{\begin{eqnarray}}
\def\ea{\end{eqnarray}}
\def\bet{\begin{tabular}}
\def\eet{\end{tabular}}

\def\pa{\partial}

\def\ve{\varepsilon}

\def\vp{\varphi}
\def\ra{\rightarrow}

\def\a{\au}
\def\m{\mu}
\def\n{\nu}
\def\s{\sigma}
\def\a{\alpha}
\def\bt{\beta}
\def\g{\gamma}
\def\dl{\delta}

\def\D{\Delta}
\def\w{\widehat}
\def\wt{\widetilde}

\def\cM{{\cal M}}
\def\cL{{\cal L}}

\def\cN{{\cal N}}
\def\cE{{\cal E}}
\def\cB{{\cal B}}
\def\cK{{\cal K}}
\def\cV{{\cal V}}

\long\def\symbolfootnote[#1]#2{\begingroup
\def\thefootnote{\fnsymbol{footnote}}\footnote[#1]{#2}\endgroup}

\begin{document}

\begin{titlepage}

\begin{flushright}
DFPD/2019/TH01\\
June 2019\\
\end{flushright}

\vspace{1truecm}

\begin{center}

{\Large \bf Duality invariant self-interactions of abelian $p$-forms in arbitrary dimensions}
\vskip1truecm

{\large Ginevra Buratti$^1$\symbolfootnote[1]{ginevra.buratti@uam.es},
Kurt Lechner$^{2,3}$\symbolfootnote[2]{kurt.lechner@pd.infn.it},
Luca Melotti$^2$\symbolfootnote[3]{luca.melotti.3@studenti.unipd.it}}

 \vspace{1.5cm}
$^1${\it Instituto de F\'isica Te\'orica UAM-CSIC, Cantoblanco, 28049 Madrid, Spain}

\vspace{1cm}
$^2${\it Dipartimento di Fisica e Astronomia ``Galileo Galilei''
\\
Universit\`a degli Studi di Padova,
Via Marzolo 8, 35131 Padova, Italy}

\vspace{1cm}
$^3${\it INFN, Sezione di Padova,
Via Marzolo 8, 35131 Padova, Italy}

\vspace{0.5cm}

\begin{abstract}

We analyze non-linear interactions of $2N$-form Maxwell fields in a space-time of dimension $D=4N$. Based on the Pasti-Sorokin-Tonin (PST) method, we derive the general consistency condition for the dynamics to respect both manifest $SO(2)$-duality invariance and manifest Lorentz invariance. For a generic dimension $D=4N$, we determine a canonical class of exact solutions of this condition, which represent a generalization of the known non-linear duality invariant Maxwell theories in $D=4$. The resulting theories are shown to be equivalent to a corresponding class of canonical theories formulated \`a la Gaillard-Zumino-Gibbons-Rasheed (GZGR), where duality is a symmetry only of the equations of motion. In dimension $D=8$, via a complete solution of the PST consistency condition, we derive new non-canonical manifestly duality invariant quartic interactions. Correspondingly, we construct new non-trivial quartic interactions also in the GZGR approach, and establish their equivalence with the former. In the presence of charged dyonic $p$-brane sources, we reveal a basic physical inequivalence of the two approaches. The power of our method resides in its universal character, reducing the construction of non-linear duality invariant Maxwell theories to a purely algebraic problem.

\end{abstract}

\end{center}

\vskip 1.0truecm \noindent {\it Keywords:} electromagnetic duality, p-form potentials, self-interactions. {\it PACS:} 11.10.Kk, 03.50.De, 11.10.Lm, 11.30.-j.
\end{titlepage}

\newpage

\baselineskip 6 mm


\tableofcontents

\section{Introduction}

Since the early years of classical electrodynamics, it is known that Maxwell's equations in empty space are manifestly invariant under continuous $SO(2)$ duality rotations of the electric and magnetic fields. At the same time, from the very beginning of the investigation of the physical consequences of this property, specifically at the quantum level \cite{DGNW,Nov}, it became clear that this simple invariance opposes a severe resistance to become a symmetry of an action: if one does not want to introduce auxiliary fields, one has either to renounce to manifest Lorentz invariance, or to a local realization of the duality transformation, or to both \cite{DT}. The uplift of an $SO(2)$ duality symmetry of the equations of motion to the level of an action becomes even more problematic for theories featuring non-linear interactions of the Maxwell fields, and/or interactions with external sources.
For an account of the implications, and relevance, of more general continuous duality symmetry groups on quantum deformations, and finiteness properties, of supersymmetric field theories and supergravity theories, see e.g. \cite{PST1,BN,EHJP} and references therein.

There are essentially three approaches to construct theories of self-interacting Maxwell fields incorporating an electromagnetic $SO(2)$ duality invariance, one of which is {\it a priori} completely unrelated to the other two. The former is due to Gaillard and Zumino \cite{GZ1,GZ2}, and has been further developed by Gibbons and Rasheed \cite{GR}; in the following it is referred to as GZGR approach. This method starts from a local, but in general non polynomial, Lagrangian $\cL(F)$ depending on a single field strength $F=dA$ of a $p$-form potential $A$, and requires that the equations of motion are invariant under $SO(2)$ rotations of the doublet $\left(F,\wt{\frac{\pa\cL}{\pa F}}\right)$, where $\sim$ denotes the Hodge dual. This requirement imposes a non-linear differential equation on $\cL(F)$, the so-called GZGR condition, which selects the duality invariant dynamics allowed for the potential $A$. One of the main advantages of this method is its manifest Lorentz invariance, its main drawback being that the Lagrangian itself is not invariant under duality rotations. Hence, in this formulation, duality is {\it not a symmetry of the action}, a feature with far-reaching consequences. For instance, it is extremely difficult to keep track of duality invariance in the quantized versions of self-interacting theories \cite{EHJP}. Moreover, when the fields are coupled to charged $p$-brane sources, in the GZGR approach this symmetry is eventually lost. In fact, the corresponding Dirac quantization condition \eref{Dirac} is not invariant under $SO(2)$ rotations of the charges, as opposed to Schwinger's quantization condition \eref{Schwinger} of the manifestly duality invariant approaches, see below. There is an important variant of the GZGR approach, due to Ivanov and Zupnik \cite{IZ1,IZ2,IZ3}, which relies on a complex formalism and suitable auxiliary fields, providing a formulation of non-linear electrodynamics in four dimensions with an inherent manifest $SO(2)$-duality invariance.
It is a peculiarity of this approach that the self-interacting parts of the actions are duality invariant, while the free parts are not. This could give rise, in principle, to an (at least technical) problem concerning the determination of the conserved Noether current associated to the $SO(2)$-invariance. Recently, the core of this approach has been extended to higher dimensions \cite{Kuz1}. A further advantage of the GZGR approach is its compatibility with  off-shell four-dimensional $N=1$ and $N=2$ supersymmetry \cite{KuTh1,KuTh2}.

The second method to construct $SO(2)$ duality invariant theories is a {\it non-covariant} first-order Hamiltonian approach,  relying on the introduction of a doublet of field-strengths $F^I=dA^I$, $I=1,2$, and of a related Lagrangian $\cL(F^I)$, which is manifestly invariant under $SO(2)$ rotations of the  potentials $A^I$. The field equations for the $A^I$ eventually reduce to a generalized first-order Hodge duality relation between, say $F^1$, and a non-linear function of $F^2$. However, being of the Hamiltonian type, this approach, introduced by Henneaux and Teitelboim \cite{HT}, and applied to quadratic duality invariant theories by Schwarz and Sen \cite{SS}, and further by Deser, Gomberoff, Henneaux and Teitelboim \cite{DGHT1,DGHT2}, is intrinsically non-manifestly covariant under Lorentz transformations: the relativistic invariance of the action, with respect to non-standard Lorentz transformation laws, must be established explicitly and leads, once more, to a differential equation on the Lagrangian $\cL(F^I)$. In a variant of this method \cite{DS,BC,BH}, the same differential equation can be derived by a criterion due to Dirac \cite{PamD} and Schwinger \cite{Schw}, requiring a  ``canonical" equal-time Poisson bracket of the energy density $T^{00}(t,\vec x)$ with itself.

A third systematic method to construct duality invariant theories has been proposed by Pasti, Sorokin and Tonin \cite{PST2,PST22}, in the following referred to as PST approach, which relies again on a doublet of field strengths $F^I=dA^I$ and, in addition, on a scalar auxiliary field $a(x)$. This approach bears the fundamental advantage of being manifestly Lorentz invariant, as well as manifestly invariant under $SO(2)$ duality, at the level of an action. A local
shift-symmetry of the field $a$ reduces the latter to a spurious, non-propagating, degree of freedom. The Lagrangian $\cL(F^I,a)$ now depends also on the field $a$, and the validity of the shift symmetry imposes on $\cL(F^I,a)$ again a differential equation, which we call the PST condition. This method has been proven to be of universal validity: it entails covariant formulations for non-linearly interacting self-dual $p$-forms \cite{PST3,PST4}, it is naturally compatible with gravity \cite{DLS}, $\kappa$-symmetry \cite{BLNPST}, with supergravity theories \cite{DLT1}, with $S$-duality \cite{DLT2}, and with the path integral quantization \cite{L,LM}. As we will further illustrate in this paper, it is also particularly efficient to analyze in a systematic manner manifestly duality invariant self-interactions of higher-rank Maxwell fields in dimensions $D=4N$. A possible flaw of the PST approach could be its conflict with off-shell supersymmetry; see, however, \cite{PST22} for a PST-off-shell version of supersymmetric chiral bosons in $D=2$.

The above approaches have been successfully applied to four-dimensional non-linear electrodynamics, and have essentially been proven to be equivalent to each other \cite{PST1,BN,DS,CKR}. More in detail, as mentioned above, in $D=4$ the duality invariant non-linear interactions are governed by specific differential equations the Lagrangians $\cL(F)$, $\cL(F^I)$ and  $\cL(F^I,a)$ must satisfy. It has been shown $i)$ that, after a suitable gauge-fixing of the auxiliary scalar field $a(x)$, typically $a(x)=x^0$, the differential equations that must be satisfied by the Lagrangians $\cL(F^I)$ and  $\cL(F^I,a)$ are the same \cite{BC}, and $ii)$ that, once these Lagrangians satisfy these equations, the resulting dynamics can be translated into the single-field formalism of GZGR, whereby the corresponding Lagrangian $\cL(F)$ satisfies automatically the GZGR condition \cite{DS}. Actually, all three differential equations are equivalent to the Courant-Hilbert equation \eref{ch2}, see \cite{CH}, particular solutions of which are the free Lagrangian $\cL(F)=-\frac{1}{4}\,F^2$, and the famous Born-Infeld Lagrangian \cite{BI,Born}.

Rather little is known about explicit examples of $SO(2)$-duality invariant self-interacting $(2N-1)$-form potentials in dimensions $D=4N$ for $D\ge 8$, apart from the basic results of the seminal paper \cite{GR}. Step forwards have been accomplished, within the GZGR approach, in references \cite{Tan,ArTan,ABMZ}, which provide a general framework for a set of $M$ self-interacting $(2N-1)$-form potentials subject to non-abelian duality groups, typically $U(M)$. However, explicit models, other than those of the Born-Infeld type, remain rare; for a review, and for some further examples of this type of duality invariant models, see reference \cite{KuTh2}.
The main difficulty in dimensions  $D\ge 8$ is that, contrary to what happens in $D=4$, generic Lagrangians depend on the field strengths $F$ and $F^I$ not only via the ``canonical" invariants $FF$ and $F\wt F$, but involve also much more complicated higher order polynomials of $F$. In particular, there is no general argument ensuring that the GZGR and PST approaches lead to physically equivalent theories. The main purpose of the present paper is to establish the basis for a systematic construction of $SO(2)$-duality-invariant theories, in the GZGR as well as in the PST formulations, and to exemplify it by deriving explicit non-trivial duality invariant interactions. Given the intrinsic complexity implied by higher-rank $p$-form potentials, the most efficient method to cope with the arising non-linearities is the PST approach. In fact, the implementation of relativistic invariance via non-canonical Lorentz transformations within the non-covariant formalisms of \cite{BN,HT,SS} and \cite{PamD,Schw} would become, in this case, rather cumbersome.

We start with a review of the GZGR formalism for $2N$-form Maxwell fields in $4N$ dimensions. We introduce the corresponding GZGR condition \eref{gzgr1} for $SO(2)$ duality invariant non-linear equations of motion, and present its {\it canonical} class of solutions in terms of the standard invariants $FF$ and $F\wt F$. Then we move on to the PST formulation, whose basic ingredient is the determination of the fundamental PST consistency condition \eref{pst22}: it guarantees that the scalar auxiliary field $a(x)$ does not propagate and, hence, that the underlying dynamics is Lorentz invariant. In fact, each (non-trivial) solution of this condition gives rise to a duality invariant self-interacting model for a Maxwell field in $D=4N$. We determine a {\it canonical} class of exact solutions also of the PST condition. Although, for canonical interactions, both the GZGR and PST conditions can formally be traced back to the same Courant-Hilbert equation \eref{ch2}, {\it a priori} it is not guaranteed that the resulting dynamics are equivalent. In particular, the former condition implies duality invariance, while the latter implies Lorentz invariance (via the decoupling of the field $a(x)$). We then give a proof, which parallels the one developed in \cite{DS} for $D=4$, that the canonical solutions of the two conditions lead, actually, to physically equivalent theories.

For dimensions $D\ge 8$, as mentioned above, there are Lorentz-invariant products of the fields $F$ (or $F^I$) that are algebraically independent of $FF$ and $F\wt F$. For $D=8$, in the search for more general solutions of the GZGR and PST conditions, we find the most general quartic invariant of the GZGR approach, which represents a deformation of the canonical theory, and, similarly, we determine the most general quartic  deformation satisfying the PST condition. Again, via a direct calculation, we verify that the deformations of the two approaches describe the same physics. This leads us to the conjecture that the procedures of, $i)$ implementing duality invariance on the equations of motion \`a la GZGR (keeping Lorentz invariance manifest) and,  $ii)$ realizing Lorentz invariance \`a la PST at the level of an action (keeping duality invariance manifest) lead always to physically equivalent theories.

Our method for constructing duality invariant self-interactions, in both approaches, is universal, in that it is reduced to merely algebraic problems, that can be resolved iteratively in the order of the self-interaction. Nevertheless, as the linear (in)dependence of Lorentz invariant polynomials of multi-index tensor fields is a non-trivial issue, in order to extend our analysis to dimensions $D\ge 12$, and to higher-order self-interactions, a more systematic procedure for solving the resulting algebraic equations would be needed. For the sake of definiteness, the analysis of the present paper is constrained to non-derivative interactions of the field strengths, although our general framework can easily be adapted to incorporate interactions of this more general kind.

According to a general argument, exemplified in \cite{Berman,Nurma} for the four-dimensional Born-Infeld electrodynamics, duality invariant self-interactions of $(2N-1)$-form Maxwell potentials in $D=4N$ should be intimately related with self-interacting chiral $2N$-form potentials in $D=4N+2$. This relation, that provides in particular a consistency check for both types of dynamics, will be investigated in a companion paper \cite{BLM}.

The paper is organized as follows. In Section \ref{gza} we review the problem of self-interacting higher-rank Maxwell fields in the GZGR formalism, with the purpose of preparing the comparison with the PST formalism. Particular attention is paid to the distinctive features, and to the equivalent (and potentially inequivalent) notions of duality, in particular to the realization of $S$-duality. We add charged dyonic sources from the beginning, as they introduce additional differences between the two approaches. In Section \ref{psta}, we rephrase the problem in the covariant, manifestly
$SO(2)$-duality invariant, PST approach, and prove the equivalence of the corresponding canonical theories. In Section \ref{quartic} we investigate the most general non-canonical quartic interactions of a non-linear Maxwell theory in $D=8$. Section \ref{out} is devoted to future perspectives.

\section{The Gaillard-Zumino-Gibbons-Rasheed approach}\label{gza}

For convenience, we write the space-time dimension as $D=2n$, with $n$ even, and introduce the $(n-1)$-form potential $A_{\m_1\cdots\m_{n-1}}$ with the associated completely antisymmetric field strength
\be\label{defeb}
F_{\m_1\cdots\m_n}=n \,\pa_{[\m_1}A_{\m_2\cdots\m_n]}.
\ee
The dynamics of this field is described by a Lorentz invariant local Lagrangian $\cL(F)$ which entails the field equations\footnote{There is an ambiguity in the normalization of the derivative with respect to antisymmetric tensor fields. Our normalization is such that for a generic variation we have the natural relation $\dl\cL=(\pa \cL/\pa F_{\m_1\cdots\m_n})\,\dl F_{\m_1\cdots\m_n}$.}
\be
\pa_{\m_1} \frac{\pa\cL}{\pa F_{\m_1\cdots\m_n}}=0.
\ee
For notational convenience, we normalize our Lagrangian in a slightly unusual way so that the Lagrangian of the free theory is\footnote{The actual Lagrangian is thus $-\cL$ or, with a more standard normalization, $-\cL/n!$.}
\be\label{lfree}
\w\cL=\frac{1}{2}\,F^{\m_1\cdots\m_n}F_{\m_1\cdots\m_n}.
\ee
For a set of $n$ completely antisymmetrized indices we introduce the compact notation
$M\equiv [\m_1\cdots\m_n]$, writing, for instance, $F_M\equiv F_{\m_1\cdots\m_n}$ and $\w\cL=\frac{1}{2}\,F^M F_M$.
For a self-interacting theory, we identify the ``dual" electromagnetic field with the Hodge dual of the tensor
\be\label{ldf}
L^{\m_1\cdots\m_n} =\frac{\pa\cL}{\pa F_{\m_1\cdots\m_n}}, \quad{\rm or}\quad L^M=\frac{\pa\cL}{\pa F_M}.
\ee
The Hodge dual is defined in a standard manner by
\[
\wt F^{\m_1\cdots\m_n} =\frac{1}{n!}\,\ve^{\m_1\cdots\m_n\n_1\cdots\n_n}F_{\n_1\cdots\n_n},\quad\quad \wt{\wt F}_M=-F_M,
\]
where we take $\ve^{01\cdots2n-1}=1$, and our Minkowski metric is $\eta^{\m\n}=(1,-1,\cdots,-1)$. Thus, Maxwell's equations for the self-interacting theory can be recast in the form
\be\label{maxi}
\begin{split}
\pa_{\m_1} \wt F^{\m_1\cdots\m_n}&=0,\\
\pa_{\m_1} L^{\m_1\cdots\m_n}&=0.\\
\end{split}\ee
In the free theory we have $L^M=F^M$.

\subsection{$SO(2)$ duality invariant equations of motion}

Inspired by the symmetric form of Maxwell's equations \eref{maxi}, we subject the doublet
$\big(F_M,\wt L_M\big)$ to an $SO(2)$ transformation by an angle $\vp\in[0,2\pi]$
\be\label{so2}
\binom{\,F^M_\vp\,}{\,\wt L^M_\vp\,}=
\left(\begin{matrix}\cos\vp & \sin\vp\\ -\sin\vp &\cos\vp\end{matrix}\right)
\left(\begin{matrix}F^M\\ \wt L^M\end{matrix}\right),
\ee
which {\it formally} keeps the form of those equations. The transformed field strength as a function of $F^M$ is thus given by
\be\label{fa}
F^M_\vp=\cos\vp \,F^M+\sin\vp\, \wt L^M.
\ee
There remains, however, the problem whether there exists a Lagrangian $\cL_\vp(\,\cdot\,)$ such that the {\it putative} transformed derivative $L^M_\vp$ of \eref{so2}
\be\label{la}
L_\vp^M=\cos\vp \,L^M +\sin\vp\, \wt F^M,
\ee
can indeed be written as
\be\label{lmfi}
L_{\vp M}=\frac{\pa\cL_\vp(F_\vp)}{\pa F_\vp^M}.
\ee
In fact, inserting the expressions \eref{fa} and \eref{la} in the right hand side of the constitutive relation
\[
d\cL_\vp(F_\vp)=dF_\vp^M L_{\vp M},
\]
a simple calculation shows that such a Lagrangian always exists, being determined implicitly by the relation
\be\label{lalpha}
\cL_\vp(F_\vp)=\cL(F)-\sin^2\vp\, FL +\frac{1}{2}\,\sin\vp\cos\vp\,\big(F\wt F+L\wt L\big).
\ee
We adopt here the convention that the unwritten indices of a product of tensors are contracted,  e.g. $FL\equiv F^M L_M$. Later on we shall use a similar notation for a partial contraction of indices
\be\label{contr}
(F^{\m_1\m_2}L^{\n_1\n_2})=F^{\m_1\m_2\rho_1\cdots \rho_{n-2}}L^{\n_1\n_2}{}_{\rho_1\cdots \rho_{n-2}},\quad \textrm{etc.}
\ee
The relation \eref{lalpha} associates to each Lagrangian $\cL(\,\cdot\,)$ a transformed Lagrangian $\cL_\vp(\,\cdot\,)$, thereby respecting, by construction, the $SO(2)$ group property. To determine $\cL_\vp(\cdot)$ operatively, one must express at the right hand side of equation \eref{lalpha} the field $F^M$ in terms of $F^M_\vp$, by inverting equation \eref{fa}, an operation that in general cannot be carried out analytically. For instance, applying the procedure to the free Lagrangian $\w\cL$ \eref{lfree} one has $L^M=F^M$, so that \eref{fa} gives $F^M=\cos\vp \,F^M_\vp-\sin\vp\, \wt F^M_\vp$. The right hand side of equation \eref{lalpha} for the transformed Lagrangian can then be computed to be $\w\cL_\vp(F_\vp)=\frac{1}{2}\,F^M_\vp F_{\vp M}$, meaning that the $SO(2)$ map leaves the free Lagrangian unaltered, $\w\cL_\vp(\,\cdot\,)=\w\cL(\,\cdot\,)$, as expected.

Introducing the fictitious Lagrangian
\be\label{kfi}
{\cal G}(F) = \cL(F)-\frac{1}{2}\,F_M\frac {\pa \cL}{\pa F_M},
\ee
the map \eref{lalpha}, which defines the transformed Lagrangian, can be recast in the alternative form
\be\label{linva}
{\cal G}_\vp(F_\vp)={\cal G}(F).
\ee
Notice that, despite the formal similarity, the Lagrangian ${\cal G}(F)$ \eref{kfi} differs from the Legendre transform of $\cL(F)$  by a factor of $1/2$. For the free Lagrangian \eref{lfree} one has  $\w{\cal G}(F)=0$.

\subsection{The GZGR condition}

Eventually, we are interested in Lagrangians that remain unaltered under a generic $SO(2)$ transformation \eref{lalpha}. We say that a Maxwell theory based on a Lagrangian $\cL$ such that
\be
\cL_\vp(\,\cdot\,) =\cL(\,\cdot\,), \quad {\rm for\,\, all}\,\,\vp,
\ee
is duality invariant. Recalling equation \eqref{lalpha}, this means that under a duality rotation the Lagrangian $\cL(F)$ itself is not invariant, but must rather obey the transformation law, first given for  $D=4$ in \cite{AFT},
\be\label{so2inv}
\cL(F_\vp)=\cL(F)-\sin^2\vp\, FL +\frac{1}{2}\,\sin\vp\cos\vp\,\big(F\wt F+L\wt L\big).
\ee
Using the equivalent definition of $\cL_\vp(\,\cdot\,)$ \eref{linva}, this law can also be expressed as the condition that the Lagrangian \eref{kfi} is invariant under the duality transformation of the field \eref{fa}
\be\label{kinv}
{\cal G}(F_\vp)={\cal G}(F).
\ee
Given the Lie-group structure of the transformation \eref{so2}, it is sufficient to satisfy equation \eref{so2inv}, or \eref{kinv}, for infinitesimal transformations. From the transformation law of the field \eref{fa} we obtain $\cL(F_\vp)=\cL(F)+\vp (L\wt L)+O(\vp^2)$, so that \eref{so2inv}, at first order in $\vp$, amounts to the {\it GZGR condition}
\be\label{gzgr1}
F\wt F-L\wt L=0.
\ee
In summary, in the GZGR formulation, a Maxwell theory is duality invariant if and only if the Lagrangian satisfies, equivalently, the conditions \eref{so2inv}, \eref{kinv}, or \eref{gzgr1}.

As in the GZGR formulation $SO(2)$ is not realized as a Noether symmetry, there is no ``canonical" local conserved current associated with it. However, if we admit a mild non-locality, then a conserved current can be constructed. In fact, the equation of motion for $A_{\m_1\cdots\m_{n-1}}$ in \eref{maxi} can be ``solved" in terms of the $S$-dual potential $A_{{\rm d}\,\m_1\cdots\m_{n-1}}$, see below, by setting
\be
\wt L_{\m_1\dots\m_n}=n\,\pa_{[\m_1}A_{{\rm d}\,\m_2\cdots\m_n]}.
\ee
A conserved current is then given by
\be\label{curgz}
j^\m=(\wt F^\m A)-(L^\m A_{\rm d}),
\ee
where we adopted the notation \eref{contr}. Using Maxwell's equations \eref{maxi}, and the GZGR condition \eref{gzgr1}, one verifies that $\pa_\m j^\m=\frac{1}{n}\,\big(\wt FF -L\wt L\big)=0$.

\subsection{$S$-duality}\label{sdual}

Although equations \eref{so2inv} and \eref{gzgr1} are equivalent formulations of the same physical condition on the dynamics of the electromagnetic field, the former entails some properties which appear rather hidden in the latter. Choosing the angle $\vp=\pi/2$, and denoting the dual field by $F_{\pi/2}^M\equiv F_{\rm d}^M$, from formula \eref{fa} and the invariance condition \eref{so2inv} we obtain
\be\label{Leg}
\cL(F_{\rm d})=\cL(F)-F_M \frac{\pa \cL}{\pa F_M}, \quad\quad F_{\rm d}^M= \wt{\frac{\pa \cL(F)}{\pa F_M}}.
\ee
These relations, in $D=4$, are sometimes interpreted by saying that in a duality invariant theory the Lagrangian is invariant under a Legendre transform. However, this is not the case, since the dual field $F_{\rm d}^M$ as a function of $F^M$ is not the derivative $\pa \cL/\pa F_M$, but rather its Hodge dual. It is, in fact, a known result that the only Lagrangians that are fixed points of the Legendre transform are the {\it free} ones $L(q)=\frac{1}{2}\,q^2$, see for instance \cite{lege}\footnote{Whether the invariance condition under the pseudo-Legendre transform \eref{Leg}, implied by the GZGR condition \eref{gzgr1}, is equivalent to the latter is, to our knowledge, not known and deserves further investigation.}. The actual content of the relations \eref{Leg} is, rather, the $S$-duality invariance of the theory, as a consequence of its $SO(2)$ duality invariance. To perform the $S$-duality transformation of the action $I[F]=\int \cL(F)\,d^Dx$ we start from the equivalent functional of $F$ and $A_{\rm d}$
\be\label{idual}
I[F,A_{\rm d}]=\int\left(\cL(F)+n\,\wt F^{\m_1\cdots\m_n}\,\pa_{\m_1}A_{{\rm d}\,\m_2\cdots\m_n}\right) d^Dx,
\ee
where $A_{\rm d}$ is the $S$-dual potential, and $F_M$ is now considered as an independent field. Varying $I[F,A_{\rm d}]$ with respect to $A_{\rm d}$, one obtains the constraint $\pa_{[\m_1}F_{\m_2\cdots\m_{n+1}]}=0 \leftrightarrow F_{\m_1\cdots\m_n}=n \,\pa_{[\m_1}A_{\m_2\cdots\m_n]}$, leading back to the original action $I[F]$. Varying $I[F,A_{\rm d}]$ instead with respect to $F$, one obtains $\wt{\frac{\pa \cL}{\pa F_M}}=n\,(\pa A_{\rm d})^M\equiv F_{\rm d}^M$. Inverting this relation to compute $F_M=F_M(F_{\rm d})$, and using this expression to eliminate from the action \eref{idual} $F$ in favor of $F_{\rm d}$, one obtains the action for the $S$-dual potential
\be\label{ifd}
I[A_{\rm d}]=\int\left(\cL(F(F_{\rm d}))- F_M \frac{\pa \cL}{\pa F_M}\right)d^Dx=\int\cL(F_{\rm d})\,d^Dx,
\ee
where in the last step we have used \eref{Leg}.
The Lagrangian for the $S$-dual field strength $F_{\rm d}$ has, thus, the same functional form as the original Lagrangian. Finally, evaluating the transformation law \eref{la} at  $\vp=\pi/2$ we obtain $L^M_{\pi/2}=\wt F^M$, i.e. (see \eref{lmfi})
\be
F^M= -\wt{\frac{\pa \cL (F_{\rm d})}{\pa F_{{\rm d}M}}}.
\ee
The same relation can be obtained by taking the derivative of the first equation in \eref{Leg} with respect to $F_{\rm d}^M$. This implies that the $S$-duality invariance relations \eref{Leg} are, in turn, equivalent to the statement that the functional relations that give $F$ in terms of $F_{\rm d}$, and $F_{\rm d}$ in terms of $F$, are the same, apart from a minus sign. To the extent that the transformation leading from \eref{idual} to \eref{ifd} can be implemented semi-classically in a functional integral approach, $S$-duality amounts to an invariance under a functional Fourier transform, rather than under a Legendre transform.

Finally, if we choose $\vp=\pi$, from \eref{fa} we obtain $F_\pi^M=-F^M$, and the $SO(2)$ invariance condition \eref{so2inv} then tells us that the Lagrangian is an even function of the field strength: $\cL(-F)=\cL(F)$. This implies that the Lagrangian of a duality invariant theory, if analytic, can contain only {\it even} powers of the field strength. This information is, actually, redundant in $D=4$, as the traces of a product of an odd number of $F^{\m\n}$'s are identically zero. This is no longer the case in dimensions $D\ge 8$, where the invariant products of an odd number of field strengths are in general non-vanishing; these products are, hence, forbidden by $SO(2)$ duality invariance.

For later reference, we introduce also the generalized electric and magnetic fields associated with the field strength $F_{\m_1\cdots\m_n}$, which are spatial antisymmetric tensors of rank $n-1$,
\be\label{ebff}
E^{i_1\cdots i_{n-1}}= F^{i_1\cdots i_{n-1}0}, \quad\quad
B^{i_1\cdots i_{n-1}}=-\frac{1}{n!}\,\ve^{i_1\cdots i_{n-1}j_1\cdots j_n}F_{j_1\cdots j_n}.
\ee
In terms of these fields we have the decompositions (to simplify the comparison with the PST approach, the spatial indices are here contracted with the Minkowski metric $\eta^{ij}=-\dl^{ij}$)
\be
\label{ffeb}
\begin{split}
FF&= n\left(E^{i_1\cdots i_{n-1}}E_{i_1\cdots i_{n-1}}-B^{i_1\cdots i_{n-1}}B_{i_1\cdots i_{n-1}}\right)=n\left(EE-BB\right),\\[5pt]
F\wt F&=2n E^{i_1\cdots i_{n-1}} B_{i_1\cdots i_{n-1}}=2nEB.
\end{split}
\ee

\subsection{Canonical theories}\label{cangzgr}

A series of analytical solutions of the GZGR condition \eref{gzgr1} can be established, if we assume that the Lagrangian depends on the field strength $F$ only via the fundamental invariants $FF$ and $F \wt F$.\footnote{In $D=4$, the products $FF$ and $F\wt F$ form a basis for {\it all} Lorentz scalars that can be formed with $F$.} For comparison with the PST approach, where Lorentz invariance holds only for a constrained class of Lagrangians, we consider here a slightly larger class of Lagrangians, namely those which are generic functions of the three independent {\it complete} contractions between $E^{i_1\cdots i_{n-1}}$ and $B^{i_1\cdots i_{n-1}}$, that is to say, $EE$, $BB$, and $EB$. We rearrange these variables in the three combinations $\{P_1,P_2,P_3\}\equiv P$
\be\label{p12}
P_1= \frac{1}{2}\,FF=\frac{n}{2}\left(EE-BB\right), \quad\quad P_2= \frac{1}{16}\,\big(F\wt F\big)^2=\frac{n^2}{4}\,(EB)^2, \quad\quad P_3=-\frac{n}{2}\,BB.
\ee
We call a theory {\it canonical}, if its Lagrangian is a function $\cL(P)$ of only these variables. In general, these Lagrangians are only invariant under the spatial rotations of $SO(D-1)$. A canonical theory is Lorentz invariant, if $\pa\cL/\pa P_3=0$. As for the GZGR condition \eref{gzgr1} for duality invariance, it translates into a differential equation for the function $\cL(P)$. Introducing the notation $\cL_i=\pa\cL/\pa P_i$, with $i=1,2,3$, a simple calculation gives
\be\label{l123}
\begin{split}
\cL_1^2-P_1\cL_1\cL_2- P_2\,\cL_2^2 + \left(\cL_1-P_3\cL_2\right)\cL_3 &=1, \quad\quad\textrm{(Duality invariance $\leftrightarrow$ GZGR condition)},\\[5pt]
\cL_3&=0,\quad\quad\textrm{(Lorentz invariance)}.
\end{split}
\ee
In particular, for a Lorentz invariant theory, the condition for duality invariance simplifies to the known form
\be\label{ch1}
\cL_1^2-P_1\cL_1\cL_2- P_2\,\cL_2^2  =1.
\ee
With the change of variables $P_1=p+q$, $P_2=-pq$, the latter can be further recast in the equivalent form
\be\label{ch2}
\cL_p\,\cL_q=1,
\ee
whose general solution has been given by Courant and Hilbert \cite{CH}. For a discussion of the physically interesting solutions of this equation, see, for instance, \cite{BC,PS,HKS}. The free Lagrangian \eref{lfree} corresponds to the solution $\cL=P_1$ of equation \eref{ch1}, and the one generalizing the four-dimensional Born-Infeld Lagrangian  $2\big(\sqrt{-\det(\eta_{\m\n}+ F_{\m\n})}-1\big)$ to dimensions $D\ge 8$ is represented by the solution\footnote{\label{scala}For dimensional reasons, in an expansion of a Lagrangian $\cL(F)$ in a power series of $F$, each term should be of the form $\lambda^N F^{N+2}$, where $\lambda$ is a coupling constant with the dimension of $({\rm length})^{D/2}$. However, since the GZGR condition \eref{gzgr1}, like all forthcoming consistency conditions, are invariant under the rescalings $F\ra kF$, $\cL\ra \cL/k^2$, it is not restrictive to set $\lambda =1$, as we shall do throughout this paper.}
\be\label{lbi}
\cL_{\rm BI}(P)=2\left(\sqrt{1+P_1-P_2}-1\right).
\ee
Actually, the general solution of equation \eref{ch1} has a universal structure up to terms of the eighth power in $F$
\be\label{lp12}
\cL(P)= P_1+ b\left(P_2+\frac{1}{4}\,P_1^2\right)  +\frac{1}{2}\,b^2P_1 \left(P_2+\frac{1}{4}\,P_1^2\right) +O(F^8),
\ee
where $b$ is a constant. The Born-Infeld Lagrangian corresponds to $b=-1$. From this formula we also see that the quartic deformation of the Lagrangian of a (Lorentz invariant) canonical theory is {\it universal}, being given by
\be\label{canquar}
b\left(P_2+\frac{1}{4}\,P_1^2\right)=\frac{b}{16}\left((F\wt F)^2+(FF)^2\right)
=\frac{bn^2}{16}\left(4(EB)^2+(BB-EE)^2 \right).
\ee
We emphasize that for dimensions $D\ge8$, contrary to what happens in a four-dimensional space-time, the polynomials \eref{p12} do by no means exhaust the algebraically independent Lorentz scalars one can form with an antisymmetric field strength $F_{\m_1\cdots\m_n}$. Therefore, in those dimensions, the solutions of the Courant-Hilbert equation \eref{ch1}, or \eref{ch2}, do not exhaust the solutions of the GZGR condition \eref{gzgr1}. This more general problem will be investigated in Section \ref{quartic}.

\subsection{Coupling to sources}\label{extsourc}

There are couplings of the electromagnetic field with external sources (or external fields), that preserve duality as a dynamical symmetry, and couplings that preserve it only at a formal level. We say that an invariance represents a dynamical symmetry of a theory, if it can be realized as a Noether symmetry of the action. Since in the GZGR formulation $SO(2)$ duality does not appear as a symmetry of this kind, it is not obvious how to distinguish couplings which maintain $SO(2)$ as a dynamical symmetry, from those which violate it. On the other hand, this distinction is straightforward in the PST formulation, where $SO(2)$ is a dynamical symmetry, see Section \ref{psta}. In general, among the interactions with external fields which preserve $SO(2)$ as a dynamical symmetry there is, for instance, the gravitational coupling to a metric $g_{\m\n}(x)$. Conversely, the coupling to charged sources violates this symmetry, keeping it only on a {\it formal} level, that is to say, the action remains invariant under $SO(2)$ rotations if one includes also a rotation of the charges. A physical consequence is that the duality current \eref{curgz} is no longer conserved, even if the charged objects become dynamical.

\subsubsection{Currents and Dirac-branes}\label{c_db}

We consider a set of $R$ charged dyonic $(n-2)$-branes, with electric charges $e^1_r$ and magnetic charges $e_r^2$, sweeping out the $(n-1)$-dimensional boundaryless world-volumes $\cM_r$, $r=1,\cdots,R$. Each world-volume is associated with a conserved current $J_r^{\m_1\cdots\m_{n-1}}(x)$ -- a tensorial $\dl$-function with support on $\cM_r$ -- satisfying $\pa_{\m_1}J_r^{\m_1\cdots\m_{n-1}}=0$. As in four-dimensional space-time, the coupling of a Maxwell field to charged dyonic sources requires, in addition, the introduction of $R$ {\it Dirac-branes} $\cN_r$, that is to say, $R$ $n$-dimensional submanifolds whose boundaries are the world-volumes $\cM_r$, $\pa \cN_r= \cM_r$. The tensorial $\dl$-function supported on the Dirac-brane $\cN_r$ is an antisymmetric tensor field $C_r^{\m_1\cdots\m_n}(x)$ of rank $n$, which by definition is related to the current by the relation\footnote{In Differential Geometry, the tensor $J_r^{\m_1\cdots\m_{n-1}}$ is the Hodge dual of the Poincar\'e-dual $(D-n+1)$-form $\Phi_r$ of the world-volume $\cM_r$, and, similarly, $C_r^{\m_1\cdots\m_n}$ is the Hodge dual of the Poincar\'e-dual $(D-n)$-form $\Psi_r$ of the Dirac-brane $\cN_r$. The relation \eref{jrdcr} is then mapped to the equality between differential forms $\Phi_r=d\Psi_r$, a basic ingredient of the notion of {\it Poincar\'e duality}.}
\be\label{jrdcr}
J_r^{\m_1\cdots\m_{n-1}}=\pa_\n C_r^{\n\m_1\cdots\m_{n-1}}.
\ee
Obviously, the Dirac-branes are not uniquely determined by the relation $\pa \cN_r= \cM_r$. If we choose a different Dirac-brane $\cN_r^*$ satisfying $\pa \cN_r^*=\cM_r$, the combined brane $\cN_r^*-\cN_r$ is, in turn, a boundaryless $n$-dimensional submanifold, and hence it is the boundary of an $(n+1)$-dimensional submanifold ${\cal S}_r$. Poincar\'e duality then implies for the associated antisymmetric tensorial $\dl$-functions the relation
\be\label{chbr}
C_r^{*\m_1\cdots\m_n}-C_r^{\m_1\cdots\m_n}=\pa_\n D_r^{\n\m_1\cdots\m_n}\quad\quad\leftrightarrow\quad\quad \cN_r^*-\cN_r=\pa{\cal S}_r,
\ee
where $D_r^{\m_1\cdots\m_{n+1}}$ is the tensorial $\dl$-function on ${\cal S}_r$.
Obviously, the choice of one Dirac-brane or another should not produce any observable physical effect.

We can now introduce the total currents and Dirac-brane $\dl$-functions ($I=1,2$)
\be\label{jc}
J^I_{\m_1\cdots\m_{n-1}} = \sum_{r=1}^Re_r^I\, J_{r\m_1\cdots\m_{n-1}},\quad\quad
C^I_{\m_1\cdots\m_n} = \sum_{r=1}^R e_r^I\, C_{r\m_1\cdots\m_n},
\ee
which are related by the constitutive relations
\be\label{jdc}
J^I_{\m_1\cdots\m_{n-1}}=\pa^\n C^I_{\n\m_1\cdots\m_{n-1}}.
\ee
The coupling of our self-interacting Maxwell field to the system of charged $(n-2)$-branes is then described by the action
\be\label{lagbr}
I[A]=\int\big(\cL(F)+n AJ^1\big)\,d^Dx, \quad \quad F_{\m_1\cdots\m_n}=n \,\pa_{[\m_1}A_{\m_2\cdots\m_n]}-\wt C^2_{\m_1\cdots\m_n},
\ee
and, accordingly, the sourceless Maxwell equations \eref{maxi} are modified to
\be\label{maxidy}
\begin{split}
\pa_{\m_1} \wt F^{\m_1\cdots\m_n}&=J^{2\m_2\cdots\m_n},\\
\pa_{\m_1} L^{\m_1\cdots\m_n}&=J^{1\m_2\cdots\m_n}.
\end{split}\ee
These equations still respect the duality invariance \eref{so2}, if we promote the charge vectors to $SO(2)$ doublets $e_r^I=(e_r^1,e_r^2)$, so that also the current vector $J^I$ in \eref{jc} transforms as an $SO(2)$ doublet\footnote{\label{fdoublet}From the transformations \eref{so2} we see that the actual doublet is $(H^1,H^2)=(F,\wt L)$. Maxwell's equations \eref{maxidy} can then be rewritten in the manifestly $SO(2)$ invariant form $\pa \wt H^I=\ve^{IJ}J^J$.}. As we see, in the Lagrangian \eref{lagbr} the couplings of the two types of charges appear in an asymmetrical fashion: the electric charges are coupled via a minimal coupling of $J^1$ to $A$, whereas the magnetic charges are coupled through the Dirac-branes $C^2$ via an additive modification of the field strength $dA$. Conversely, as we will see in the next Section, in the manifestly duality invariant PST approach both types of charges will be coupled in both ways: via a minimal coupling {\it and} via a Dirac-brane.

If the Lagrangian $\cL$ satisfies the GZGR condition \eref{so2inv}, or \eref{gzgr1}, the  $S$-duality transformation of the action \eref{lagbr} can again be performed explicitly, see Section \ref{sdual}. This time we find, in place of \eref{ifd},
\be\label{Isad}
I[A_{\rm d}]=\int\big(\cL(F_{\rm d})+n A_{\rm d} J^2- C^1\wt C^2\big)\,d^Dx, \quad\quad F_{{\rm d}\,\m_1\cdots\m_n}=n\,\pa_{[\m_1}A_{{\rm d}\,\m_2\cdots\m_n]}+
\wt C^1_{\m_1\cdots\m_n}.
\ee
As in the sourceless case, the non-linear kinetic term of the dual field strength equals the original one, but the electric and magnetic charges have interchanged their role according to the replacements $C^1\ra C^2$, $C^2\ra- C^1$, or $e^1_r\ra e_r^2$, $e_r^2\ra- e_r^1$. This transformation corresponds to the generator
\be\label{z4}
\left(\begin{matrix}0 & 1\\ -1 &0\end{matrix}\right)
\ee
of the discrete subgroup $Z_4$ of $SO(2)$, corresponding to the rotation angle $\vp=\pi/2$, which represents, indeed, $S$-duality.

\subsubsection{Charge quantization}\label{chquant}

The $S$-dual action \eref{Isad} contains, actually, the unexpected additional term $-\int C^1\wt C^2\, d^Dx$, built with the $\dl$-functions supported on the Dirac-branes, which at first sight does not admit a simple interpretation. However, it becomes innocuous if we resort to Dirac's quantization condition, inherent in every physical system of charged objects endowed with electric as well as magnetic charges. In fact, although the action \eref{lagbr} formally exhibits a dependence on the choice of Dirac-branes $\cN_r$, this dependence becomes spurious if the charges $e_r^I$ satisfy the Dirac quantization condition. The argument is standard, but as there will emerge a basic difference with the manifestly $SO(2)$ invariant PST approach, we sketch it here briefly.

If we make the change of Dirac-branes $\cN_r\ra\cN_r^*$, the corresponding $\dl$-functions change according to the rule \eref{chbr}. If we want to keep the field strength $F_{\m_1\cdots\m_n}$ in \eref{lagbr} unaltered -- an observable quantity -- then the gauge field $A$  must change by a combination of the Hodge duals of the tensors $D_r$, involving the magnetic charges $e_r^2$,
\be\label{Achange}
A_{\m_1\cdots\m_{n-1}}\ra A_{\m_1\cdots\m_{n-1}} + \sum_r e_r^2\,\wt D_{r\m_1\cdots\m_{n-1}}.
\ee
Under this transformation, the minimal-interaction term produces in the action \eref{lagbr} the ``Dirac anomaly"
\be\label{changei}
I[A]\ra I[A]+n\sum_{r,s}e_r^2e_s^1 \int  \wt D_r  J_s\,d^Dx.
\ee
However, according to Poincar\'e duality, the integrals appearing in the Dirac anomaly just count the number of intersections $N_{rs}$ between the $(n-1)$-dimensional submanifold occupied by the $s$-th $(n-2)$-brane, and the $(n+1)$-dimensional submanifold ${\cal S}_r$ involved in the change of the $r$-th $(n-2)$-brane, see for instance \cite{LM1}.  More precisely, we have $\int  \wt D_r J_s\,d^Dx=(n-1)! N_{rs}$. Recalling that the actual action is given by $I[A]/n!$, we see that the exponentiated action $\exp\left(\frac{i}{n!}\, I[A]\right)$ does not depend on the choice of Dirac-branes, whenever the charges satisfy the {\it Dirac quantization conditions}
\be\label{Dirac}
e_r^2e_s^1=2\pi n_{rs}, \quad n_{rs}\in {\mathbb Z},\quad\forall r,s.
\ee
As anticipated, these conditions do not respect $SO(2)$ duality. On the other hand, under the discrete $Z_4$ transformation \eref{z4} the product of charges appearing in \eref{Dirac} changes by
\be
e_r^2e_s^1 \ra -e_r^1e_s^2.
\ee
Although not invariant, if the original product is an integer multiple of $2\pi$, so is the final one, in compatibility with Dirac's condition \eref{Dirac}. This signals that, in the GZGR formulation, in the presence of sources quantum effects break the continuous $SO(2)$ symmetry of the equations of motion down to its discrete subgroup $Z_4$ (see \cite{LM} for a concrete realization of the symmetry $Z_4$ in the context of a relativistic quantum field theory of dyons in four dimensions). Returning to the additional term in the $S$-dual action \eref{Isad}, this term can now be rewritten as the double sum
\be
\int C^1\wt C^2\, d^Dx= \sum_{r,s}e_r^1e_s^2 \int C_r\wt C_s\,d^Dx.
\ee
This time, the integral counts the number of intersections $M_{rs}$ of the Dirac-branes as $\int C_r\wt C_s\,d^Dx=n! M_{rs}$. If the charges satisfy the quantization condition \eref{Dirac}, then this term drops out from the exponential $\exp\left(\frac{i}{n!}\,I[A_{\rm d}]\right)$, and so is physically irrelevant.

Finally, the couplings appearing in the action \eref{lagbr} can also be traded for interactions with external fields, as they typically occur in type IIA and type IIB supergravity theories. In this case, the $n$-forms $C^I$ turn into Chern-Simons forms constructed with products of other $p$-form potentials, or into spinor bilinears, or into a sum of such terms.

\section{The manifestly $SO(2)$ invariant PST approach}\label{psta}

We investigate now the dynamics of a non-linear Maxwell theory in a two-potential formulation. For notational convenience, we write the dimension as $D=2n=2p+2$, where the integer $p=n-1$ is {\it odd}. For this purpose, we introduce a doublet of $p$-form potentials $A^I_{\m_1\cdots\m_p}$, with $I=1,2$. Their {\it magnetic} coupling to the dyonic sources introduced in Section \ref{extsourc}, see equations \eref{jrdcr}-\eref{jdc}, is then realized via the modified field strengths
\be\label{fai}
F^I_{\m_1\cdots\m_n}=n \,\pa_{[\m_1}A^I_{\m_2\cdots\m_n]}- \wt C^I_{\m_1\cdots\m_n},
\ee
which satisfy the Bianchi identities
\be\label{bianchi}
\pa_{\m_1} \wt F^{I\m_1\cdots\m_n}=J^{I\m_2\cdots\m_n}.
\ee
Aim of this Section is to generalize the non-manifestly covariant approaches of the two-potential formulations of references \cite{SS,DS}, to determine all possible manifestly $SO(2)$-duality invariant self-interactions of $n$-form Maxwell-fields, thereby keeping also Lorentz invariance manifest. As anticipated in the Introduction, the most efficient method for the realization of this program is represented by the PST approach \cite{PST1,PST2}.

The key ingredient in this formalism is a scalar auxiliary field $a(x)$, whose gradient $\pa_\m a$ we suppose to be a non-vanishing time-like vector. Then we can introduce the unit vector
\[
v^\m=\frac{\pa^\m a}{\sqrt{(\pa a)^2}},\quad\quad v^2=1,
\]
which allows to decompose a generic antisymmetric tensor in a component along $v^\m$, and a component orthogonal to it. We exemplify the procedure for the fields we are interested in, namely the field strengths \eref{fai}. For this purpose, we define their parallel components $\cE^I$, and their orthogonal components $\cB^I$, both tensors of rank $p$, as
\be\label{be}
{\cE}^I_{\m_1\cdots\m_p}= F^I_{\m_1\cdots\m_p\n}v^\n,\quad\quad {\cB}^I_{\m_1\cdots\m_p}=\wt F^I_{\m_1\cdots\m_p\n}v^\n.
\ee
Then, the field strengths can be decomposed as
\be\label{1}
F_{\m_1\cdots\m_n}^I=n\,{\cE}_{[\m_1\cdots\m_{n-1}}^Iv_{\m_n]}-
\frac{1}{(n-1)!}\,\ve_{\m_1\cdots\m_n}{}^{\n_1\cdots\n_n} {\cB}_{\n_1\cdots\n_{n-1}}^Iv_{\n_n}.
\ee
The symbols $\cE^I$ and $\cB^I$ have not been chosen by chance, for if we set $v^\m=\dl^{\m0}$, then the tensors \eref{be} have only spatial components, which reduce precisely to the electric and magnetic fields \eref{ebff} associated with an antisymmetric Maxwell tensor.
There are some useful identities regarding the invariant squares one can form with the field strengths $F^I$ (the contracted indices of biproducts are suppressed also in this section)
\be\label{2}
F^IF^J=n\left(\cE^I \cE^J-\cB^I\cB^J\right)=-\wt F^I\wt F^J,\quad\quad F^I\wt F^J = n\left(\cE^I\cB^J+\cE^J\cB^I\right),
\ee
to be compared with the relations \eref{ffeb}.
Integrating the second decomposition in \eref{2} over the whole space-time, inserting the expressions for the field strengths  \eref{fai}, and using the relations for the currents \eref{jdc}, we obtain the further useful identity
\be\label{identeb}
\int\left(\cE^I\cB^J+\cE^J\cB^I\right)d^Dx = -\int\left(A^IJ^J+A^JJ^I\right)d^Dx -\frac{1}{n}\int C^I \wt C^J\,d^Dx.
\ee

\subsection{Covariant action and PST condition}

The action we propose is a functional of the doublet $A^I$ and of the auxiliary field $a$, whereas the sources $J_r^I$, or $C_r^I$, are considered as external fields. It has the manifestly $SO(2)$ invariant form
\be
\label{saa}
S[A,a]=-n\int\left(\frac{1}{2}\,\varepsilon^{IJ}\left(\cE^I \cB^J+A^IJ^J\right) +\cN(\cB)\right)d^Dx,
\ee
where the {\it Hamiltonian} $\cN(\cB)$ is a Lorentz and $SO(2)$ invariant function of the magnetic fields only. In fact, $\cN(\cB)$ will become the {\it true} Hamiltonian after the elimination of, say, the gauge field $A^1$ in favour of $A^2$, see Section \ref{gzgrpst}. This action can be also written in an alternative, non-manifestly $SO(2)$ invariant form, using the identity \eref{identeb} for $I=1$ and $J=2$,
\be\label{saanon}
S[A,a]=n\int\left(\cE^2 \cB^1+A^2J^1 -\cN(\cB)\right)d^Dx +\frac{1}{2}\int C^1\wt C^2\,d^Dx.
\ee
For the choice $v^\m=\dl^{\m0}$, and in the absence of sources, the action $S[A,a]$ reduces $i)$ for $\cN\big(\cB\big)= \frac{1}{2}\,\cB^I\cB^I$ to the non-covariant linear theories in $D=2n$ of reference \cite{SS}, and $ii)$ for $D=4$ to the non-covariant interacting theories of reference \cite{DS}. The overall coefficient has been chosen to match with the normalization of the GZGR action \eref{lfree}, see below, while the relative coefficient between the kinetic terms $\cE\cB$ and the
minimal-coupling terms $AJ$ is fixed by the peculiar PST-symmetries of the action, to be discussed in a moment.

As for every theory of bosonic fields, the action \eref{saa} is of second order in the derivatives and, as it stands, propagates two gauge potentials. To retrieve the correct number of degrees of freedom, the resulting second-order equations of motion should reduce {\it spontaneously} to first-order differential equations, which eventually should assume the form of a non-linear Hodge-duality relation between the two field strengths; this would then trace the dynamics back to a single Maxwell field. The other peculiar aspect of the action \eref{saa} is the presence of the auxiliary field $a$, which eventually should not represent a physical degree of freedom. Both these goals -- halving the degrees of freedom of the Maxwell fields, and eliminating the auxiliary field $a$ -- are achieved by means of the PST symmetries.

\subsubsection{Symmetries and equations of motion}

The action \eref{saa} is trivially invariant under the standard gauge transformations
$\dl A^I_{\m_1\cdots\m_p}=\pa_{[\m_1}\Lambda^I_{\m_2\cdots\m_p]}$. To establish the occurrence of additional local symmetries of the action we first determine its variation
\be\label{totvar}
\delta S=\delta S\big|_{\dl A}+\delta S\big|_{\dl a}
\ee
under arbitrary variations of the potentials and of the auxiliary field. Under a change $A^I\ra A^I+\dl A^I$ the action varies as
\be\label{dAi}
\delta S\big|_{\dl A}=
\frac{n}{p!}\int\ve^{\m_1\cdots\m_p\n_1\cdots\n_p\rho\s}\ve^{IJ}\pa_\s
\left(h^I_{\m_1\cdots\m_p}v_\rho\right)\dl A^J_{\n_1\cdots\n_p}d^Dx,
\ee
while for $a\ra a+\dl a$ one finds
\be\label{da}
\delta S\big|_{\dl a}=
-\frac{n}{2p!}\int\!\ve^{\m_1\cdots\m_p\n_1\cdots\n_p\rho\s}v_\rho\dl v_\s
\left(\ve^{IJ}\cB^I_{\m_1\cdots\m_p}\cB^J_{\n_1\cdots\n_p}+ \ve^{IJ}\cE^I_{\m_1\cdots\m_p} \cE^J_{\n_1\cdots\n_p}+2N^I_{\m_1\cdots\m_p} \cE^I_{\n_1\cdots\n_p}\right)\!d^Dx.
\ee
The variation of the vector $v^\m$ is $\dl v^\m= (\eta^{\m\n}-v^\m v^\n)\,\pa_\n\dl a/\sqrt{(\pa a)^2}$, but the second term drops out from the right hand side of equation \eref{da}. Above, we have introduced the rank-$p$ tensor doublets
\be\label{defnh}
h^I_{\m_1\cdots\m_p}=\cE^I_{\m_1\cdots\m_p}+\ve^{IJ} N^J_{\m_1\cdots\m_p},\quad\quad
N^I_{\m_1\cdots\m_p}= \frac{\pa \cN(\cB)}{\pa\cB^{I\m_1\cdots\m_p}}.
\ee
Notice that the variation $\ve^{IJ}\dl A^I J^J$ canceled out against a similar term arising from the variation of the kinetic term $\ve^{IJ}\cE^I \cB^J$ of the action \eref{saa}. This compensation imposes, indeed, both the presence and the normalization of the minimal-interaction term in \eref{saa}. Vice versa, the presence of the minimal-interaction term would force the presence of the Poincar\'e-duals $\wt C^I$ of the Dirac-branes in the field strengths \eref{fai}.

From the form of the above variations it is immediately seen that the action is invariant under the transformations (PST1)
\be\label{PST1}
\delta A^I_{\m_1\cdots\m_p}=\pa_{[\m_1}a\,\lambda^I_{\m_2\cdots\m_p]},\quad\quad \dl a=0,
\ee
where  $\lambda^I_{\m_1\cdots\m_{p-1}}$ is an antisymmetric transformation parameter of rank $p-1$. In this case we have trivially $\delta S\big|_{\dl a}=0$, whereas the vanishing of $\delta S\big|_{\dl A}$ follows via an integration by parts. Notice that the transformations \eref{PST1} are a symmetry of the action \eref{saa}, independently of the functional form of $\cN(\cB)$. There is a further, more involved, {\it potential} symmetry of this action, that shifts the auxiliary field $a$ by an arbitrary scalar field $\Phi$. It has the form (PST2)
\be\label{PST2}
\delta A^I_{\m_1\cdots\m_p}=-\frac{\Phi}{\sqrt{(\pa a)^2}}\,
h^I_{\m_1\cdots\m_p},\quad\quad
\delta a=\Phi.
\ee
Under these transformations the total variation of the action \eref{totvar} combines into the compact expression
\be
\dl S=-\frac{n}{2p!}\int\ve^{\m_1\cdots\m_p\n_1\cdots\n_p\rho\s}v_\rho \dl v_\s\,\ve^{IJ}\left(
\cB^I_{\m_1\cdots\m_p}\,\cB ^J_{\n_1\cdots\n_p}- N^I_{\m_1\cdots\m_p}\,
N^J_{\n_1\cdots\n_p}\right) d^Dx.
\ee
Therefore, the scalar field $a(x)$ becomes a spurious degree of freedom, if the Hamiltonian $\cN(\cB)$ satisfies the {\it PST condition} (generalizing analogous conditions found previously for four-dimensional space-times \cite{PST1,BN})
\be\label{pst22}
\cB^I_{[\m_1\cdots\m_p}\,\cB ^J_{\n_1\cdots\n_p]}- N^I_{[\m_1\cdots\m_p}\,
N^J_{\n_1\cdots\n_p]}=0.
\ee
In fact, by construction the magnetic fields $\cB^I$ and the tensors $N^I$ have no components along the direction $v^\m$, i.e. $v^{\m_1}\cB^I_{\m_1\cdots\m_p}=0=v^{\m_1}N^I_{\m_1\cdots\m_p}$. In the PST approach, once the field $a(x)$ has been gauge-fixed, the constraint \eref{pst22} represents eventually the condition for Lorentz invariance, whereas $SO(2)$ invariance is manifest. Conversely, in the GZGR approach, the constraint \eref{gzgr1} represents the condition for $SO(2)$ invariance, whereas Lorentz invariance is manifest.

If the condition \eref{pst22} holds, the equation of motion of the scalar field takes a particularly simple form. In fact, enforcing \eref{pst22}, and recalling the definition of the tensors $h^I$ \eref{defnh}, from the general variations of the action \eref{dAi} and \eref{da} we derive the field equations for $A^I$ and $a$, respectively,
\begin{align}
\pa_{[\m_1}\left(v_{\m_2} h^I_{\m_3\cdots\m_{p+2}]}\right)&=0,\label{eqmai}\\[5pt]
\ve^{IJ}\ve^{\rho\s\m_1\cdots\m_p\n_1\cdots\n_p}\,\pa_\rho\bigg(\frac{\pa_\s a}{(\pa a)^2}\,
h^I_{\m_1\cdots\m_p}\,h^J_{\n_1\cdots\n_p}\bigg)&=0.\label{eqma}
\end{align}
In equation \eref{eqmai} the antisymmetrization over all Lorentz indices is understood. As the field $a$ is a pure-gauge degree of freedom, its equation of motion should not add any new dynamical constraint on the system. In fact, it is easily seen that equation \eref{eqma} is a mere consequence of the equations of motion of the gauge potentials \eref{eqmai}, which are hence the only ones we have to cope with. Due to the Poincar\'e lemma, the general solution of these equations can be written in the form
\be\label{vh}
v_{[\m_1}h^I_{\m_2\cdots\m_{p+1}]}=\pa_{[\m_1}a\,\pa_{\m_2}\Lambda^I_{\m_3\cdots\m_{p+1]}},
\ee
for some doublet of tensor fields $\Lambda^I_{\m_1\cdots\m_{p-1}}$. On the other hand, under a PST1 transformation \eref{PST1} we have
\be
\dl\left(v_{[\m_1}h^I_{\m_2\cdots\m_{p+1}]}\right)=
\pa_{[\m_1}a\,\pa_{\m_2}\lambda^I_{\m_3\cdots\m_{p+1]}}.
\ee
Therefore, if we choose the gauge parameters $\lambda^I_{\m_1\cdots\m_{p-1}}=-\Lambda^I_{\m_1\cdots\m_{p-1}}$, equations \eref{vh} simply reduce to $v_{[\m_1}h^I_{\m_2\cdots\m_{p+1}]}=0$. But, since the tensors $h^I_{\m_1\cdots\m_p}$ \eref{defnh} have by definition no components along $v^\m$,  $v^{\m_1} h^I_{\m_1\cdots\m_p}=0$, the latter equations are in turn equivalent to the generalized self-duality relations
\be\label{genselfd}
h^I_{\m_1\cdots\m_p}=\cE^I_{\m_1\cdots\m_p}+\ve^{IJ} N^J_{\m_1\cdots\m_p}(\cB)=0.
\ee
Equations \eref{genselfd} represent the desired first-order gauge-fixed version of the second-order equations of motion \eref{eqmai} of the Maxwell fields.

\vskip0.5truecm
\noindent
{\it Conserved $SO(2)$-duality current.} If the Hamiltonian $\cN(\cB)$ is an $SO(2)$ invariant function of the fields $\cB^I$, i.e. $\dl\cN(\cB)=\vp\,\ve^{IJ}\cB^J N^I=0$, and if we take vanishing sources, $J^I=C^I=0$, then the action \eref{saa} is invariant under the Noether symmetry $\dl A^I=\vp \ve^{IJ}A^J$. The associated conserved current is easily calculated to be
\[
j^\m=\frac{1}{n!}\,\ve^{\m\m_1\cdots\m_p\n_1\cdots\n_{p+1}}\left(F^I_{\n_1\cdots\n_{p+1}}
-2n\, h^I_{\n_1\cdots\n_p}v_{\n_{p+1}}\right)A^I_{\m_1\cdots\m_p},\quad\quad \pa_\m j^\m=0.
\]
In the PST1-gauge where $h^I=0$, it reduces to the simpler expression
\be\label{jfixed}
j^\m= \wt F^{I\m\m_1\cdots\m_p} A^I_{\m_1\cdots\m_p}.
\ee
It matches with the corresponding current \eref{curgz} of the GZGR approach, if we set $A^2=A$, $A^1=-A_{\rm d}$, and enforce the identifications \eref{f12lm} below.
Using the identities in \eref{2}, and the gauge-fixed equation of motion \eref{genselfd}, the divergence of $j^\m$ can be computed to be
\[
\pa_\m j^\m= \frac{1}{n}\,\wt F^I F^I=2\cB^I\cE^I=-2\cB^I\ve^{IJ}N^J,
\]
which indeed vanishes thanks to the $SO(2)$ invariance of $\cN(\cB)$.
The conserved $SO(2)$-duality charge is thus given by
\be\label{dualq}
Q=\int  \wt F^{I0\m_1\cdots\m_p}A^I_{\m_1\cdots\m_p}\,d^{D-1}x.
\ee
In four-dimensional space-time, the expression \eref{jfixed} reduces to the known duality current, see for instance \cite{AF,ARN}. If, in $D=4$, we consider the linear theory, then the quantity $Q/\hbar$ can be seen to be equal to the difference of the numbers of  photons with positive and negative helicity, see for instance \cite{ARN}. For a non-linear theory in a generic $D$-dimensional space-time the physical interpretation of the duality charge \eref{dualq} still needs to be investigated.

\vskip0.5truecm
\noindent
{\it $S$-duality.} As we did for the GZGR formulation in Section \ref{sdual}, we now test the properties of the action \eref{saa} under an $S$-duality transformation. The procedure is standard: we consider the field strengths \eref{fai} as independent fields $F^I_{\m_1\cdots\m_n}$, and impose the Bianchi identities \eref{bianchi} via a Lagrange multiplier doublet $ \w A^I_{\m_1\cdots\m_{n-1}}$, which represent the $S$-dual potentials. The $S$-dual field strengths are hence given by
\[
\w F^I_{\m_1\cdots\m_n}=n \,\pa_{[\m_1}\w A^I_{\m_2\cdots\m_n]}- \wt C^I_{\m_1\cdots\m_n}.
\]
Then the action  \eref{saa} is equivalent to the action
\be\label{saasdual}
S[F,\w A,a]=-n\int\left(\frac{1}{2}\,\varepsilon^{IJ}\left(\cE^I \cB^J -\frac{1}{n}\,\wt F^I\w F^J + \w A^IJ^J  \right) +\cN(\cB)\right)d^Dx.
\ee
In fact, varying $S[F,\w A,a]$ with respect to $\w A^I$ we obtain the Bianchi identities \eref{bianchi}, with solution \eref{fai}. Substituting the latter back in \eref{saasdual} we recover the original action \eref{saa}. On the other hand, from the action \eref{saasdual} we can now eliminate the field $F^I$ in favor of $\w A^I$. The equation of motion of the former is
\be
F^I_{\m_1\cdots\m_n}=\w F^I_{\m_1\cdots\m_n}+2n
h_{[\m_1\cdots\m_{n-1}}^Iv_{\m_n]}.
\ee
Introducing for the field strengths $\w F^I$ electric and magnetic fields $\w\cE^I$ and $\w\cB^I$ as in  \eref{be}, these equations of motion amount to $\cE^I=\w\cE^I+2h^I$, $\cB^I=\w\cB^I$. Substituting the latter back in \eref{saasdual}, and using that $\wt F^I\w F^J=n\big(\cE^I\w\cB^J+\cB^I\w\cE^J\big)$, see \eref{2}, we get the action for the $S$-dual potentials
\be
S\big[\w A,a\big]=-n\int\left(\frac{1}{2}\,\varepsilon^{IJ}\left(\w\cE^I\, \w\cB^J+ \w A^IJ^J  \right) +\cN\big(\w\cB\big)\right)d^Dx,
\ee
which has precisely the same form as the original action \eref{saa}, irrespective of the functional form of the Hamiltonian $\cN(\cB)$. The PST formalism bears thus an (essentially manifest) $S$-duality invariance, staying on the same footing of its manifest $SO(2)$ duality invariance. Observe that this is not what happens in the GZGR formulation, where $S$-duality is a symmetry of the action only if $\cL(F)$ satisfies the GZGR condition \eref{gzgr1}.

\subsubsection{Charge quantization}

Again we may ask to which extent the action \eref{saa} depends on the choice of Dirac-branes. Extending the notations of Section \ref{chquant} to both Maxwell potentials $A^I$, to keep the field strengths \eref{fai} invariant under a change of Dirac-branes, the potentials must transform as in \eref{Achange}
\be\label{AchangeI}
A^I_{\m_1\cdots\m_p}\ra A^I_{\m_1\cdots\m_p} + \sum_r e_r^I\,\wt D_{r\m_1\cdots\m_p}.
\ee
This time, the minimal-interaction term of the action \eref{saa} leads to the Dirac anomaly
\be\label{pstch}
S[A,a]\ra S[A,a]-\frac{n}{2}\,\sum_{r,s}\ve^{IJ} e_r^I\,e_s^J \int  \wt D_r  J_s\,d^Dx,
\ee
to be compared with the Dirac anomaly \eref{changei} of the GZGR approach. The invariance of $\exp(iS[A,a]/n!)$ then requires that the charges satisfy the {\it Schwinger quantization conditions}\footnote{For dyonic $p$-branes in dimensions $D\ge 4$, within the framework of linear theories, these conditions have first been derived in \cite{DGHT1,DGHT2}, where, however, the factor $1/2$ at the left hand side of equations \eref{Schwinger} -- a purely relativistic effect -- was missed.}
\be\label{Schwinger}
\frac{1}{2}\left(\,e_r^2e_s^1-e_r^1e_s^2\,\right)=2\pi n_{rs}, \quad n_{rs}\in {\mathbb Z},\quad\forall r,s,
\ee
to be compared with the Dirac conditions \eref{Dirac}. There emerges thus a basic difference between the GZGR and PST formulations of duality invariant (non-linear) Maxwell theories, when coupled to sources: the PST formulation is manifestly invariant under $SO(2)$ rotations and so is the corresponding quantization condition \eref{Schwinger}, which guarantees the unobservability of the Dirac-branes. Conversely, in the GZGR formulation $SO(2)$ is not a symmetry of the action and so the corresponding quantization condition \eref{Dirac} is not constrained by this symmetry. Therefore, in the presence of sources, we cannot expect the two formulations to be equivalent; actually, they will not be so, as we will see in the next section. The difference between the quantization conditions \eref{Dirac} and \eref{Schwinger} has important physical consequences. In particular, in $D=4$, the GZGR approch entails a phenomenon known as {\it spin-statistics transmutation}: it turns the $r$-th dyon from a boson into a fermion, and vice versa, if the integer $e_r^1e_r^2/2\pi$ is ${\it odd}$, see for instance \cite{LM2}. No such phenomenon occurs in the PST approach, where Schwinger's quantization condition \eref{Schwinger} does not require any relation between the charges $e_r^1$ and $e_r^2$ of the $r$-th $(n-2)$-brane.

\subsection{Relation with the GZGR approach}\label{gzgrpst}

As mentioned earlier, there is no stringent {\it a priori} reason for which the GZGR and PST approaches should be physically equivalent; rather, as in $D=4$, this equivalence must be established case by case. Before facing the issue of the equivalence, we must determine the precise relation between the two formulations. There are essentially two possible ways to establish such a relation in practice.

\vskip0.5truecm\noindent
{\it Method I.} The first method requires to invert the generalized self-duality conditions \eref{genselfd} to express the electric and magnetic fields of one field strength in terms of those of the other, say
\be\label{edib}
\cE^1 =\cE^1(\cE^2,\cB^2),\quad\quad \cB^1 =\cB^1(\cE^2,\cB^2).
\ee
Recalling the decompositions \eref{be}, \eref{1}, these relations allow to express the field strength $F^1$ in terms of $F^2$ and of the auxiliary field $a$,
\be\label{f12a}
F^1=F^1(F^2,a).
\ee
However, as the action \eref{saa} is invariant under the PST2 transformations \eref{PST2}, so must be the equations of motion. Now, since we have already fixed the PST1 transformations according to \eref{genselfd}, i.e. $h^I=0$, the former simply reduce to $\dl a=\Phi$, $\dl A^I=0$. This implies that in the (Lorentz invariant) relations \eref{f12a} there is, actually, no dependence on $a$! As a last step, one should be able to rewrite these relations in the form
\be\label{f1f2}
\wt F^1_{\m_1\cdots\m_n}=\frac{\pa\cL(F^2)}{\pa F^{2\m_1\cdots\m_n}},
\ee
for some Lagrangian $\cL(F^2)$. Indeed, the second method ensures that this is always possible, see below. Then, via the identifications
\be\label{f12lm}
  F_M^1=-\wt L_M,\quad\quad F_M^2=F_M,
\ee
the Bianchi identities \eref{bianchi} go indeed over to the GZGR Maxwell equations \eref{maxidy}\footnote{The doublet $(F^1,F^2)=(-\wt L,F)$ is related to the GZGR doublet $(H^1,H^2)=(F,\wt L)$ in \eref{so2} by the $SO(2)$-covariant relation $H^I=\ve^{IJ}F^J$, see Footnote \ref{fdoublet} in Section \ref{c_db}. Notice also that $F^1$ coincides with minus the $S$-dual $F_{\rm d}$ of the field strength $F$, see \eref{Leg}.}. After that, there remains the problem to analyze the interrelation between the PST condition \eref{pst22} and the GZGR condition \eref{gzgr1}, in general a non-trivial task, see Section \ref{eqth} for the particular case of the {\it canonical} theories.

\vskip0.5truecm\noindent
{\it Method II.} The second method to relate the two formulations aims to reconstruct the GZGR action \eref{lagbr} from the PST action \eref{saa}, by eliminating, say, the potential $A^1$. Correspondingly, we enforce the (gauge-fixed) equation of motion \eref{genselfd} only for the potential $A^1$, namely $h^2=0$, or
\be\label{eqa1}
\cE^2=\frac{\pa\cN(\cB^1,\cB^2)}{\pa\cB^1},
\ee
and we invert it to determine $\cB^1= \cB^1(\cE^2,\cB^2)$. Substituting this relation in the PST action \eref{saa}, or better, in its equivalent form \eref{saanon}, we see that the resulting action becomes a functional of only $A^2$ of the form
\be\label{sa22}
S[A^2]=\int\big(\cL(F^2)+n A^2J^1\big)\,d^Dx + \frac{1}{2}\int C^1\wt C^2\,d^Dx,
\ee
where the Lagrangian
\be\label{lagf2}
\cL(F^2)=n\left(\cE^2\cB^1 -\cN(\cB^1,\cB^2)\right)\Big|_{\cB^1=\cB^1(\cE^2,\cB^2)},
\ee
would {\it a priori} also depend on the auxiliary field $a$. But
again, the action \eref{sa22} still inherits those PST symmetries that have not been gauge-fixed. To obtain the equation of motion $h^2=0$ we have fixed the PST1 symmetry \eref{PST1} for $A^2$, which is thus no longer a symmetry of the action \eref{sa22}. More importantly, the PST2 symmetry survives. However, since we have that $h^2=0$, the PST2 transformations \eref{PST2} simply reduce to $\dl a= \Phi$, $\dl A^2=0$.\footnote{If $h^2=0$, the equation of motion of $a$ \eref{eqma} is satisfied identically and, correspondingly, the action \eref{sa22} is invariant under an arbitrary shift $a\ra a+\Phi$. Recall, however, that the equation of motion of $a$ takes the remarkable form \eref{eqma}, only thanks to the PST condition \eref{pst22}. It is this condition that eventually ensures the decoupling of $a$, and hence Lorentz invariance.} This implies that the -- manifestly Lorentz invariant -- Lagrangian  \eref{lagf2} does not depend on $a$, and hence it is invariant under ``standard'' Lorentz transformations of $A^2$.

In light of relation \eref{eqa1}, equation \eref{lagf2} defines $\cL(F)$ as the {\it Legendre transform} of the {\it Hamiltonian} $\cN(\cB)$, and vice versa. In particular, from the generalized duality relations -- equation \eref{f1f2} in the GZGR formulation, and equations \eref{genselfd} in the PST formulation -- we derive the corresponding symplectic differential identities
\be\label{simplec}
d\cL=n\left(d\cE^2\cB^1+d\cB^2\cE^1\right),\quad\quad d\cN=d\cB^1\cE^2-d\cB^2\cE^1.
\ee
Via the identification $A^2=A$, the action \eref{sa22} matches with the GZGR action $I[A]$ \eref{lagbr}, modulo the last term. This term represents a non-trivial self-interaction of the Dirac-branes, and signals the inequivalence of the two formulations in the presence of sources:
\be\label{azgzpst}
S[A^2]=I[A]+\frac{1}{2}\int C^1\wt C^2\,d^Dx.
\ee
Notice that under a change of Dirac-branes \eref{chbr} this term changes by
\be
\frac{1}{2}\int C^1\wt C^2\,d^Dx\ra \frac{1}{2}\int C^1\wt C^2\,d^Dx -\frac{n}{2}
\sum_{r,s}e_r^2e_s^1 \int\left(\wt D_r  J_s+\wt D_s J_r\right)d^Dx.
\ee
If we add this variation to the Dirac anomaly \eref{changei} of the GZGR action $I[A]$, we recover the correct $SO(2)$-invariant Dirac anomaly of the PST action \eref{pstch}.

\vskip0.5truecm\noindent
{\it Effective actions.} The relation \eref{azgzpst} between the PST and GZGR actions suggests a simple modification of the latter, which could induce a symmetry enhancement of the original GZGR action from $Z_4$ to $SO(2)$, at least at the level of the effective action of the currents.
In fact, if we replace the GZGR action $I[A]$ \eref{lagbr} with
$I[A]+\frac{1}{2}\int C^1\wt C^2\,d^Dx$,  which leaves the equations of motion \eref{maxidy} unaltered, then the above argument shows that this modified GZGR action entails the $SO(2)$ invariant Dirac anomaly \eref{pstch}, leading to Schwinger's quantization condition \eref{Schwinger}. This is, of course, not enough to ensure the full $SO(2)$ symmetry of the (low energy) quantum theory. For this to be the case, the effective action $\Gamma_{\rm GZGR}[J]$ for the currents $J=\{J^1,J^2\}$ should be $SO(2)$ invariant. The latter is defined by the functional integral (putting aside ultraviolet divergences, requiring a cut-off, and the gauge fixing)
\[
e^{i\Gamma_{\rm GZGR}[J]}=\int\{{\cal D}A\}\,e^{i\left(I[A]+\frac{1}{2}\int C^1\wt C^2\,d^Dx\right)}.
\]
The functional $\Gamma_{\rm GZGR}[J]$ entails the same $SO(2)$ invariant Dirac anomaly \eref{pstch} of the modified GZGR action, because the measure $\{{\cal D}A\}$ is invariant under the translation \eref{Achange}. But one might now expect that $\Gamma_{\rm GZGR}[J]$ itself is $SO(2)$ invariant, thanks to the duality invariance condition \eref{so2inv}, or \eref{gzgr1}, satisfied by the GZGR Lagrangian $\cL(F)$. This invariance property has indeed been proven for a generic linear Maxwell theory in dimensions $D=2n$ in \cite{LM,LM1}, where the functional $\Gamma_{\rm GZGR}[J]$ has been evaluated explicitly, whereas for non-linear theories the problem remains open. In the PST approach, the effective action $\Gamma_{\rm PST}[J]$ is obtained from the PST action $S[A,a]$ \eref{saa} by the double functional integral (no functional integration over $a$ is required, as the effective action is already $a$-independent)
\[
e^{i\Gamma_{\rm PST}[J]}=\int\{{\cal D}A^1\}\{{\cal D}A^2\}\,e^{iS[A,a]}.
\]
$\Gamma_{\rm PST}[J]$ is trivially invariant under $SO(2)$ rotations of the currents, because $S[A,a]$ is invariant under joined $SO(2)$ rotations of $A^I$ and $J^I$, and it carries again the Dirac anomaly \eref{pstch}. The final challenge would be the comparison of the effective actions
$\Gamma_{\rm GZGR}[J]$ and $\Gamma_{\rm PST}[J]$, for a GZGR Lagrangian  $\cL(F)$ tied to the PST Hamiltonian $\cN(\cB)$ by the Legendre transform \eref{lagf2}. As this relation is of purely classical origin, it is problematic to argue that the two quantum effective actions are the same, although they carry the same Dirac anomaly and are (presumably) both invariant under $SO(2)$.

In conclusion, the PST formalism for a duality invariant non-linear Maxwell theory always allows to construct an equivalent, manifestly Lorentz invariant, theory in terms of a single field strength $F=dA$, and a related Lagrangian $\cL(F)$, as foreseen by the GZGR approach. However, it is not guaranteed that $\cL(F)$ satisfies the GZGR condition \eref{gzgr1}, which would ensure the $SO(2)$ duality invariance of the resulting equations of motion. The path can also be inverted: starting from a Lorentz invariant Lagrangian $\cL(F)$ satisfying the GZGR condition, the Legendre transform \eref{lagf2} reconstructs a Hamiltonian $\cN(\cB)$ that satisfies automatically the PST condition, but it is not guaranteed that this Hamiltonian is invariant under $SO(2)$. This issue will be settled explicitly in Section \ref{eqth} for the case of {\it canonical} theories.

\subsubsection{Linear theory and deformations}\label{lindef}

We now exemplify the above methods for the linear theory, where both can be carried out analytically. This theory is represented by the quadratic Hamiltonian $\cN(\cB)= \frac{1}{2}\,\cB^I\cB^I$, which satisfies the PST condition \eref{pst22} trivially. In this case we have in fact the simple relation $ N^I_{\m_1\cdots\m_p}={\cB}^I_{\m_1\cdots\m_p}$,
and the generalized self-duality relations \eref{genselfd} become
\be\label{hii}
h^I_{\m_1\cdots\m_p}= {\cE}^I_{\m_1\cdots\m_p}+\ve^{IJ}{\cB}^J_{\m_1\cdots\m_p}=0.
\ee
According to the first method, we have to put these relations in the form $F^1=F^1(F^2)$. For this purpose, we introduce the doublet of rank-$n$ antisymmetric tensors $H^I=F^I+\ve^{IJ}\wt F^J$, which are tied to the tensor \eref{hii} by the relation $h^I_{\m_1\cdots\m_p}= H^I_{\m_1\cdots\m_p\n}v^\n$, recall the definitions \eref{be}. $H^I$ obeys the twisted self-duality identity $\wt H^I=-\ve^{IJ}H^J$. Decomposing this doublet in the same manner as the field strengths in \eref{1},
\be\label{hij}
H_{\m_1\cdots\m_n}^I=n\,h_{[\m_1\cdots\m_{n-1}}^Iv_{\m_n]}+
\frac{1}{(n-1)!}\,\ve^{IJ}\,\ve_{\m_1\cdots\m_n}{}^{\n_1\cdots\n_n} h_{\n_1\cdots\n_{n-1}}^Jv_{\n_n},
\ee
we see that equations \eref{hii} are equivalent to the manifestly covariant relations $H^I=0$, i.e.
\be
\wt F^1_{\m_1\cdots\m_n}= F^2_{\m_1\cdots\m_n}.
\ee
Comparing with the general formula \eref{f1f2}, we have thus retrieved the quadratic Lagrangian $\cL(F^2)=\frac{1}{2}\,F^2F^2$ of the GZGR formulation. Proceeding with the second method, we enforce the equation of motion \eref{hii} for $A^1$
\[
h^2=\cE^2-\cB^1=0.
\]
Inserting the Hamiltonian of the linear theory $\cN(\cB)= \frac{1}{2}\,\cB^I\cB^I$, and replacing $\cB^1$ with $\cE^2$, the Lagrangian \eref{lagf2} then becomes, see \eref{2},
\be\label{laglin}
\cL(F^2)=n\left(\cE^2\cB^1 - \frac{1}{2}\left(\cB^1\cB^1+\cB^2\cB^2\right) \right)
=\frac{n}{2}\left(\cE^2\cE^2-\cB^2\cB^2\right)=\frac{1}{2}\,F^2F^2,
\ee
which is again the Lagrangian of the linear theory.

Moving away from the linear theory, we observe that the Hamiltonian $\cN(\cB)$ of a generic non-linear theory can be always split up into the quadratic term of the linear theory, and a deformation
\be\label{splitn}
\cN({\cB})=\frac{1}{2}\,\cB^I\cB^I+ \cM(\cB).
\ee
Due to $SO(2)$ invariance, the deformation $\cM(\cB)\equiv \cM(\cB^1,\cB^2)$ can contain only {\it even} powers of the $\cB^I$, and so it starts with quartic powers of these fields. A similar split-up can be made for the GZGR Lagrangian
\be\label{splitl}
\cL(F)=\frac{1}{2}\,FF+\cK(F).
\ee
Again, the deformation $\cK(F)$ starts with quartic powers of $F$, since a duality invariant Lagrangian $\cL(F)$ must be an even function of $F$ (see the end of Section \ref{sdual}). This time, the relation \eref{eqa1} is more complicated
\be\label{e2b1}
\cE^2=\cB^1+ \frac{\pa\cM(\cB^1,\cB^2)}{\pa \cB^1} \quad \leftrightarrow \quad \cB^1=\cE^2+\cV(\cE^2,\cB^2),
\ee
where the so-defined tensor $\cV_{\m_1\cdots\m_p}(\cE^2,\cB^2)$ starts with cubic powers of $\cB^2$ and $\cE^2$. Inserting the magnetic  field $\cB^1$ \eref{e2b1}, and the decomposition of the Hamiltonian \eref{splitn}, in the general formula for the GZGR Lagrangian \eref{lagf2}, we see that the terms linear in the deformation $\cV(\cE^2,\cB^2)$ drop out, and we obtain
\be
\cL(F^2)=\frac{n}{2}\left(\cE^2\cE^2-\cB^2\cB^2\right)-n\cM(\cE^2,\cB^2)+O^6(\cE^2,\cB^2).
\ee
A comparison with the decomposition \eref{splitl} then yields that, at quartic order, the GZGR  deformation $\cK(F)$ is related to the PST deformation $\cM(\cB)$ by the simple formula
\be\label{karel}
\cK(F^2)=-n\,\cM(\cE^2,\cB^2)+ O\left((F^2)^6\right).
\ee
This implies, in particular, that there is a {\it universal} simple relation between the {\it quartic} interactions of the two formulations, which we will investigate in Section \ref{quartic}. According to the general characteristics of Method $II$, formula \eref{karel} also shows that, once we replace $\cB^1$ with $\cE^2$ in the quartic term $\cM(\cB)\big|_4$ of the deformation $\cM(\cB)$, and the former satisfies the PST condition \eref{pst22}, we automatically obtain a {\it Lorentz invariant} quartic polynomial $\cK(F)\big|_4$ in $F^2= F$. However, even restricting ourselves to quartic deformations, and despite the simple relation \eref{karel}, there does not seem to exist a simple argument showing the equivalence between the GZGR and PST approaches, i.e. an argument proving that, once $\cM(\cB)\big|_4$ is an $SO(2)$-invariant polynomial, then $\cK(F)\big|_4$ satisfies the GZGR duality condition \eref{gzgr1}, and vice versa.

\subsection{Canonical theories}\label{canthpst}

Parallel to the GZGR formulation, see Section \ref{cangzgr}, there is a class of theories for which the Hamiltonian $\cN(\cB)$ depends on the fields $\cB^I$ only via their {\it complete} contractions, namely the three scalar products $\cB^1\cB^1$, $\cB^2\cB^2$ and $\cB^1\cB^2$. We call a theory of this kind {\it canonical}. If we insist on $SO(2)$ duality invariance, these three independent invariants collapse,  actually, into two: the quadratic invariant $\cB^I\cB^I$, and the quartic invariant $(\cB^I\cB^J)(\cB^I\cB^J)$. For this reason, it is convenient to rearrange the three scalar products above in the combinations $\{Q_1,Q_2,Q_3\}=Q$
\be\label{q123}
Q_1=\frac{1}{2}\,\cB^I\cB^I,\quad\quad  Q_2=\frac{1}{8}\left((\cB^I\cB^J)(\cB^I\cB^J)
-(\cB^I\cB^I)^2\right),\quad\quad Q_3 = \frac{1}{2}\,\cB^2\cB^2.
\ee
Accordingly, we write the Hamiltonian of a canonical theory as $\cN(Q)$\footnote{\label{wd4}In $D=4$, the polynomials $Q_1$ and $Q_2$ are the unique independent Lorentz- and duality-invariant polynomials one can form with the vectors $\cB^I_\m$. In four dimensions, the invariant $Q_2$ is usually written as the square $W^\m W_\m$ of the vector \cite{DS,Nurma}
\be
W^\m=\ve^{IJ}\ve^{\m\s\,\m_1\cdots\m_p\,\n_1\cdots\n_p}
\cB^I_{\m_1\cdots\m_p}\cB^J_{\n_1\cdots\n_p}v_\s,
\ee
which resembles, in a sense, the invariant $(F\wt F)^2$ of the GZGR formulation. In fact,
in $D=4$, the usual formula for the product of two Levi-Civita tensors yields for this square the simple expression $W^\m W_\m=16 Q_2$. In contrast, for $D\ge8$ the same calculation yields for $W^\m W_\m$ a sum of a variety of quartic polynomials, see equation \eref{ww} in $D=8$, which would not allow for simple solutions of the PST condition \eref{pst22}.}. Correspondingly, the theory is duality invariant, if $\pa\cN/\pa Q_3=0$.
Introducing the notation $\cN_i=\pa\cN/\pa Q_i$, for this particular functional dependence the PST condition \eref{pst22} can be evaluated to be
\be\label{pstcan}
\begin{split}
\cB^I_{[\m_1\cdots\m_p}\,\cB^J_{\n_1\cdots\n_p]}-& N^I_{[\m_1\cdots\m_p}\,N^J_{\n_1\cdots\n_p]}=\\[10pt]
\big\{1-\cN_1^2+Q_1\cN_1\cN_2 +Q_2\cN_2^2-&\left(\cN_1-Q_3\cN_2
\right)\cN_3\big\}\,\cB^I_{[\m_1\cdots\m_p}\,\cB^J_{\n_1\cdots\n_p]}.
\end{split}
\ee
Since equation \eref{pst22} ensures the decoupling of $a$ or, equivalently, the relativistic invariance of the underlying theory, we so derive the conditions
\be\label{n123}
\begin{split}
\cN_1^2 -Q_1\cN_1\cN_2-Q_2\cN_2^2+\left(\cN_1-Q_3\cN_2\right)\cN_3 &=1, \quad\quad\textrm{(Lorentz invariance $\leftrightarrow$ PST condition)},\\[5pt]
\cN_3&=0,\quad\quad\textrm{(Duality invariance)}.
\end{split}
\ee
For a duality invariant theory, $\cN_3=0$, the first equation reduces to
\be\label{ch3}
\cN_1^2-Q_1\cN_1\cN_2- Q_2\cN_2^2=1,
\ee
which is formally identical to the GZGR condition  \eref{ch1}, and so, via an appropriate change of variables, it can again be reduced to the Courant-Hilbert equation \eref{ch2}. As happens in the GZGR formulation, up to the eighth power of $\cB^I$ the solution of equation \eref{ch3} is universal, see the expansion \eref{lp12},
\be\label{nq12}
\cN(\cB)= Q_1+ c \left(Q_2+\frac{1}{4}\,Q_1^2\right)   +\frac{1}{2}\,c^2Q_1\left(Q_2+\frac{1}{4}\,Q_1^2\right)+O(\cB^8).
\ee
The universal quartic deformation in a canonical theory is thus given by
\be\label{canquarpst}
c\left(Q_2+\frac{1}{4}\,Q_1^2\right)=\frac{c}{8}\left((\cB^I\cB^J)(\cB^I\cB^J)
-\frac{1}{2}\,(\cB^I\cB^I)^2\right)=
 \frac{c}{16}
\left(4(\cB^1\cB^2)^2+\left(\cB^1\cB^1-\cB^2\cB^2\right)^2
\right).
\ee
There is a surprising formal coincidence of the equations appearing in \eref{l123} and \eref{n123}. However, the role of the conditions is flipped: the equation that ensures Lorentz invariance on the PST side ensures duality invariance on the GZGR side, and vice versa. We did not find any deeper reason, or interpretation, of this ``duality", which remains thus mysterious\footnote{In $D=4$, and for $\cN_3=0=\cL_3$, the fact that with an appropriate choice of quadratic and quartic canonical invariants the Lorentz invariance condition on the PST side, and the duality condition on the GZGR side, are formally identical, has first been noted in \cite{BC}. The authors of this reference ascribe this coincidence to a ``natural" property of the Legendre transform \eref{lagf2}, without furnishing, however, any motivation.}.

\subsubsection{Linking the PST and GZGR formulations}\label{eqth}

If we choose $v^\m=(1,0,\cdots,0)$, the non-vanishing components of the magnetic fields are the spatial components $\cB^I_{i_1\cdots i_{n-1}}$. Then, as in the polynomials \eref{q123} all indices are pairwise contracted, the map \eref{lagf2} associates to a generic canonical Hamiltonian $\cN(Q)$ a generic canonical Lagrangian $\cL(P)$, depending on all three invariants $P_1$, $P_2$ and $P_3$, see \eref{p12}. As we saw in Section \ref{gzgrpst}, $\cN(\cB)$ satisfies the PST condition \eref{pst22}, if and only if $\cL(F)$ is Lorentz invariant. For canonical theories, this implies that $\cN(Q)$ satisfies the differential equation in \eref{n123}, if and only if $\cL_3=0$, see \eref{l123}, i.e. if $\cL(P)$ does not depend on $P_3$. This is the first point we want to check explicitly in this section. The second point we want to prove is, instead, a non-trivial result: $\cL(P)$ satisfies GZGR condition in \eref{l123}, if and only if $\cN_3=0$, i.e. if $\cN(Q)$ does not depend on $Q_3$. In other words, $\cL(P)$ is duality invariant, if and only if $\cN(Q)$ is so. As the relation \eref{lagf2} between $\cL(P)$ and $\cN(Q)$ is only implicit, the calculations are rather cumbersome and we relegate them to Appendix \ref{link}. The results of this section represent, in particular, a generalization to generic dimensions $D\ge 8$ of a seminal analysis performed for vector potentials in $D=4$ in reference \cite{DS}.

The main result of Appendix \ref{link} are the relations between the derivatives of $\cN(Q)$ and $\cL(P)$ (due to the identification $F=F^2$, we have that $P_3=-nQ_3$, see the formulas in \eref{p12} and \eref{q123})
\be\label{nxyzP}
\begin{split}
\cN_1&=\frac{1}{\cL_1+nQ_3\cL_2},\\[10pt]
\cN_2&=-\frac{n\cL_2/\cL_1}{\cL_1+nQ_3\cL_2},\\[10pt]
\cN_3&=\frac{\cL_1^2-P_1\cL_1\cL_2-P_2\cL_2^2+\left(\cL_1-P_3\cL_2\right)\cL_3-1}{\cL_1+nQ_3\cL_2},
\end{split}
\ee
where $Q_1$ and $Q_2$ can, in turn, be expressed in terms of $P_1$, $P_2$ and $P_3$, see  formulas \eref{defx} and \eref{defy}. As shown in Appendix \ref{link}, using these expressions one can prove the following two basic identities
\begin{align}
\label{n1233}&\cN_1^2 -Q_1\cN_1\cN_2-Q_2\cN_2^2+\left(\cN_1-Q_3\cN_2\right)\cN_3-1=\frac{\cL_3}{\cL_1},\\[10pt]
\label{equivfin}&\frac{\cL_1^2-P_1\cL_1\cL_2- P_2\cL_2^2-1}{\cL_1}=-\frac
{\cN_1^2-Q_1\cN_1\cN_2- Q_2\cN_2^2-1}{\cN_1}.
\end{align}
The identity \eref{n1233} provides a check of the first point mentioned above, i.e. the equivalence between the PST condition in \eref{n123}, and the Lorentz invariance of the GZGR formulation. The third relation in \eref{nxyzP} proves, instead, the second point, namely that the PST Hamiltonian is duality invariant, $\cN_3=0$, if and only if the related equations of motion of the GZGR theory are so, see \eref{l123}. Finally, the identity \eref{equivfin} proves the double implication, holding for Lorentz {\it and} duality invariant theories,
\be\label{ln12}
\cL_1^2-P_1\cL_1\cL_2-P_2\cL_2^2=1\quad\quad\Leftrightarrow\quad\quad\cN_1^2-Q_1\cN_1\cN_2- Q_2\cN_2^2 =1.
\ee
In other words, the Legendre transformation \eref{lagf2} associates to each solution $\cN(Q_1,Q_2)$ of equation \eref{ch3} a solution $\cL(P_1,P_2)$ of equation \eref{ch1}, and vice versa. This means that -- in the case of  canonical theories -- a duality invariant Hamiltonian $\cN(\cB)$ satisfying the PST condition \eref{pst22} is associated with a unique Lorentz invariant Lagrangian $\cL(F)$ satisfying the GZGR condition \eref{gzgr1}, and vice versa.
We conjecture that this important relation holds also for generic theories -- not depending solely on canonical invariants -- as we shall verify explicitly for generic quartic interactions in $D=8$ in Section \ref{quartic}.

In general, the map $\cL(P)\leftrightarrow \cN(Q)$ is difficult to realize explicitly. Apart from the quadratic theory, it is easy to determine explicitly the Hamiltonian $\cN_{\rm BI}(Q)$ associated to the Born-Infeld Lagrangian $\cL_{\rm BI}(P)$ \eref{lbi}. As $\cL_{\rm BI}(P)$ satisfies the first differential equation in \eref{ln12}, it is guaranteed that $\cN_{\rm BI}(Q)$ satisfies the second differential equation. In fact, in this case we have $\cL_2=-\cL_1$, and so the relations \eref{nxyzP} imply  $\cN_2=n\cN_1$, with $\cN_1>0$ for small fields. The second equation in \eref{ln12} then reduces to $\cN_1=1/\sqrt{1-nQ_1-n^2Q_2}$, yielding\footnote{The solutions of the differential equations \eref{ln12} are scale invariant, see Footnote \ref{scala} of Section \ref{cangzgr}.}
\be\label{bipst}
\cN_{\rm BI}(Q)=\frac{2}{n}\left(1-\sqrt{1-nQ_1 -n^2Q_2}\,\right).
\ee
More in general, as the canonical quartic interactions of both the GZGR and PST formulations are universal, they must go over into each other. Indeed, from their explicit expressions \eref{canquar} and \eref{canquarpst}, respectively, we see that they fit with the matching condition \eref{karel}, if we set $c=-nb$: operating the replacement $\cB^1\ra\cE^2$, the deformation $-n\cM(\cB)$ of \eref{canquarpst} goes over into the Lorentz invariant deformation $\cK(F)$ \eref{canquar}.

\section{Duality invariant quartic interactions in $D=8$}\label{quartic}

It seems rather difficult to solve the PST condition \eref{pst22} analytically, if we allow the Hamiltonian $\cN(\cB)$ to be a generic $SO(2)$-duality invariant (and formally Lorentz invariant) function of the fields $\cB^I$, as it is difficult to solve the GZGR condition  \eref{gzgr1} for a generic Lorentz invariant Lagrangian $\cL(F)$. In this section, we face this problem for a self-interacting Maxwell theory in an eight-dimensional space-time, with the aim of determining the most general quartic interactions satisfying these conditions. As in this case we have no {\it a priori} indication of the equivalence of the PST and GZGR approaches, concerning duality, one of our purposes will be to investigate their interrelation.

\subsection{Quartic deformations in the GZGR formulation}

As we are only interested in quartic deformations,  we rewrite the GZGR condition \eref{gzgr1} as a condition on the deformation $\cK(F)$ introduced in equation \eref{splitl}, i.e. $\cL(F)=\frac{1}{2}\,FF+\cK(F)$. Denoting  the derivatives of $\cK(F)$ with respect to the field strength by
\[
K^{\m_1\cdots\m_4}=\frac{\pa\cK(F)}{\pa F_{\m_1\cdots\m_4}}\equiv K^M,
\]
the GZGR condition $F \wt F = L \wt L$ translates into the differential equation for $\cK(F)$
\be\label{gzgrk}
\wt F K=-\frac{1}{2}\,\wt K K.
\ee
As $\cK(F)$ starts with terms of order $F^4$, the left hand side of this equation is of order $F^4$, while its right hand side starts with terms of order $F^6$. Therefore, understanding henceforth with $\cK(F)$ the {\it quartic} deformation of  the theory, equation \eref{gzgrk} reduces to the linear equation $\wt FK=0$, which can be read as an invariance condition for $\cK(F)$:
\be\label{invgz}
\delta \cK(F)=0, \quad\quad \dl F=\wt F.
\ee
The main purpose of this section is to determine the most general Lorentz invariant solution $\cK(F)$ of this equation, which represents, hence, the most general duality invariant quartic deformation within the GZGR approach.

We begin by writing down a basis for the independent Lorentz invariant quartic polynomials one can form with the field strength $F_{\m_1\cdots\m_4}$. There are four of them\footnote{Here we assume the theory to be parity preserving, so that no Levi-Civita tensor can appear in $\cK$.}
\begin{align}
\cK_1&=(FF)^2,\label{k1}\\[5pt]
\cK_2&=(F^\m F^\n)(F_\m F_\n),\label{k2}\\[5pt]
\cK_3&=(F^{\m\n} F^{\rho\s})(F_{\m\n} F_{\rho\s}),\label{k3}\\[5pt]
\cK_4&=(F^{\m\n} F^{\rho\s})(F_{\m\rho} F_{\n\s})\label{k4},
\end{align}
where again we understand that the non-written indices are contracted.
Now we have to solve the algebraic problem of determining the combinations of these polynomials that satisfy the invariance condition \eref{invgz}, alias the GZGR condition. Although the involved tensor algebra is conceptually simple, there are some hidden relations inherent in it, expressed by the following identities. The first identity is, actually, straightforward and introduces just an alternative way of parameterizing the above basis
\be\label{fwf2}
(F\wt F)^2=-\frac{8!}{4!4!}\,F^{[\m_1\cdots\m_4}F^{\m_5\cdots\m_8]}F_{\m_1\cdots\m_4}
F_{\m_5\cdots\m_8} =-2\cK_1+32\cK_2-36 \cK_3.
\ee
The other two identities we need involve one Levi-Civita tensor
\begin{align}
\label{simple}\big(F^\m \wt F^\n\big)& =\frac{1}{8}\,\eta^{\m\n}F\wt F,\\[5pt]
\label{compl}\big(F^{\m\n} F^{\rho\s}\big)\big(F_{\m\n} \wt F_{\rho\s}\big)&=\frac{1}{12}\,(FF)(F\wt F).
\end{align}
To prove the identity \eref{simple}, write out the product
\be\label{simplep}
\big(F^\m \wt F^\n\big)=F^{\m\rho_1\rho_2\rho_3}\wt F^\n{}_{\rho_1\rho_2\rho_3},
\ee
and then write $F^{\m\rho_1\rho_2\rho_3}$ as minus the Levi-Civita tensor times its Hodge dual, and insert for $\wt F^\n{}_{\rho_1\rho_2\rho_3}$ its definition in terms of $F$. Using the standard formula giving the product of two Levi-Civita tensors in terms of multiple products of Minkowski metrics, the right hand side of \eref{simplep} becomes
\be
\big(F^\m \wt F^\n\big)= \frac{1}{4}\,\eta^{\m\n} F\wt F -\big(F^\m \wt F^\n\big),
\ee
which is \eref{simple}. The proof of the identity \eref{compl} is more involved and is outlined in Appendix \ref{idencom}, see below for an indirect derivation.

It is now straightforward to compute the variations under $\dl F=\wt F$ of the polynomials \eref{k1}-\eref{k4}
\begin{align}
\dl\cK_1&=4(FF)(F\wt F),\label{dk1}\\[5pt]
\dl\cK_2&=\frac{1}{2}\,(FF)(F\wt F),\label{dk2}\\[5pt]
\dl\cK_3&= \frac{1}{3}\,(FF)(F\wt F),\label{dk3}\\[5pt]
\dl\cK_4&= 4(F^{\m\n} F^{\rho\s})(F_{\m\rho}\wt F_{\n\s}) \neq c(FF)(F\wt F).\label{dk4}
\end{align}
The variation \eref{dk1} is trivial. The variation \eref{dk2} follows from the identity \eref{simple}, and \eref{dk3} follows from \eref{compl}. A consistency check of these variations is provided by the identity \eref{fwf2}. In fact, the variation
$\dl(F\wt F)^2=-4(FF)(F\wt F)$ agrees with formulas \eref{dk1}-\eref{dk3}. This provides the independent proof of the identity \eref{compl}, mentioned above. The inequality \eref{dk4}, implying that $\cK_4$ cannot appear in a duality invariant polynomial, is proven in Appendix \ref{idencom}.

We thus see that, out of the four polynomials \eref{k1}-\eref{k4}, there are precisely two indipendent combinations which satisfy the GZGR condition \eqref{invgz}, namely
\be\label{inv1}
\cK_3-\frac{1}{12}\,\cK_1, \quad\quad\cK_2 - \frac{1}{8}\,\cK_1.
\ee
Using the decomposition \eref{fwf2}, they can be rearranged as
\be\label{inv2}
R_1=(F\wt F)^2+\cK_1, \quad\quad R_2=\cK_2 - \frac{1}{8}\,\cK_1.
\ee
As expected, one of the polynomials, namely $R_1$, is the universal canonical quartic invariant \eref{canquar}. The invariant $R_2$ represents instead a new, non-canonical, duality invariant quartic interaction.

\subsection{Quartic deformations in the PST formulation}\label{quartpst}

As in the GZGR formulation, we rewrite the PST condition \eqref{pst22} on the Hamiltonian \eqref{splitn} $\cN(\cB)=\frac{1}{2}\cB^I \cB^I + \cM(\cB)$ as a condition on the deformation $\cM(\cB)$. Denoting its derivatives by
\be
M^{I \m_1\cdots \m_p}=\frac{\pa \cM(\cB)}{\pa \cB^I_{\m_1 \cdots \m_p}},
\ee
this time we obtain the differential equation for $\cM(\cB)$
\be\label{pstalin}
\cB^{[I}_{[\m_1\cdots\m_p}\,M^{J]}_{\n_1\cdots\n_p]}=-\frac{1}{2}\,  M^I_{[\m_1\cdots\m_p}\,M^J_{\n_1\cdots\n_p]},
\ee
in place of \eref{gzgrk}. As before, since $\cM(\cB)$ starts with quartic powers of the fields $\cB^I$, the left hand side of this equation starts with terms of order four, while its right hand side starts with terms of order six. Therefore, the PST condition for the quartic deformation of the theory, that we continue to denote with $\cM$, reads
\be\label{pstlin}
\cB^{[I}_{[\m_1\cdots\m_p}\,M^{J]}_{\n_1\cdots\n_p]}=0.
\ee
To rewrite this condition in a more convenient form, we introduce a ``transformation parameter" $\Delta_\mu$, and define the formal variation
\be \label{deltab}
\dl \cB^{I \n_1 \cdots \n_p}=\frac{1}{p!}\,\D_\s\,\ve^{IJ} \ve^{\rho \s \m_1 \cdots \m_p \n_1 \cdots \n_p}\cB^J_{\m_1 \cdots \m_p} v_\rho.
\ee
Then equation \eref{pstlin} can be recast equivalently as the invariance condition for $\cM$ under the transformations \eref{deltab} (recall that the fields $\cB^I_{\m_1 \cdots \m_p}$ are orthogonal to $v^\m$)
\be \label{deltaa}
 \D_\s\,\ve^{IJ} \ve^{\rho \s \m_1 \cdots\m_p \n_1 \cdots\n_p} \cB^J_{\m_1 \cdots \m_p} v_\rho M^I_{\n_1 \cdots \n_p}=0 \quad \leftrightarrow \quad \dl \cM(\cB)=0,
\ee
which parallels the GZGR condition \eref{invgz}.

Let us now search for solutions of this equation in an eight-dimensional space-time, where the magnetic fields are a doublet of three-tensors $\cB^{I\m_1\m_2\m_3}$. For this purpose, it is convenient to perform an $SO(1,7)$ Lorentz transformation to rotate the unit vector $v^\m$ to the simple form $v^\m=(1,0,\cdots,0)$. This has the advantage that the only non-vanishing components of the fields are the spatial ones $\cB^{In_1 n_2 n_3}$, $n=(1,\dots,7)$. We also introduce the seven-dimensional Hodge dual of $\cB^{In_1 n_2 n_3}$ (in the remainder of this section the spatial indices will be contracted with the euclidean metric $\dl^{mn}$)
\be
\wt \cB^{In_1\cdots n_4}= \frac{1}{3!}\,\ve^{n_1\cdots n_7}\cB^{In_5 n_6 n_7},\quad\quad
\cB^{In_1 n_2 n_3}= \frac{1}{4!}\,\ve^{n_1\cdots n_7}\wt \cB^{In_4\cdots n_7}.
\ee
In this frame, the variation \eqref{deltab} simplifies to
\be \label{deltab2}
\dl \cB^{I n_1n_2n_3}=-\D^m \ve^{IJ} \wt{\cB}^{Jmn_1n_2n_3}.
\ee
Before facing the solution of the PST condition $\dl \cM(\cB)=0$, one can ask how many independent quartic solutions are expected to exist. As we know from Section \ref{gzgrpst}, this condition gives rise on the GZGR side to Lorentz invariant quartic polynomials, and there are precisely four of them, see \eref{k1}-\eref{k4}. In addition, we are interested in $SO(2)$ invariant polynomials, and, as we will see, this further constraint reduces the number of independent solutions from four to {\it two}. To determine them explicitly, it is more convenient to impose first $SO(2)$ invariance, and only then the PST condition.

A basis of $SO(2)$ invariant and Lorentz invariant (in this form, actually, $SO(7)$ invariant) quartic polinomials of $\cB^{In_1n_2n_3}$ is given by the six elements\footnote{To write these polynomials in $SO(1,7)$ invariant form, it suffices to replace $m\ra \m$, $n\ra\n$, etc., and to contract the indices with the Minkowski metric.}
\begin{align}
\cM_1&=(\cB^I \cB^I)^2,\label{a1} \\[5pt]
\cM_2&=(\cB^{I m} \cB^{I n})(\cB^{Jm} \cB^{Jn}), \\[5pt]
\cM_3&= (\cB^{I m} \cB^{J n})(\cB^{Im} \cB^{Jn}), \\[5pt]
\cM_4&= (\cB^{I m} \cB^{J n})(\cB^{In} \cB^{Jm}), \\[5pt]
\cM_5&=(\cB^I \cB^J)(\cB^I \cB^J), \\[5pt]
\cM_6&=(\cB^{I mn} \cB^{I kl})(\cB^{Jmk} \cB^{Jnl}), \label{a6}
\end{align}
where, as usual, it is understood that the non-written indices are contracted. The polynomials $\cM_1$ and $\cM_5$ rephrase the canonical invariants $Q_1$ and $Q_2$ in \eref{q123}. To impose the PST condition \eref{deltaa}, we must compute the variations of these polynomials under a transformation \eqref{deltab2}
\begin{align}
\dl \cM_1&=-4 \ve^{IL} \D^m (\cB^J \cB^J)(\cB^I \wt{\cB}^{L m}), \label{dlA1} \\[5pt]
\dl \cM_2&=-\frac{2}{3}\,\ve^{IL} \D^m \left((\cB^J \cB^J)(\cB^I \wt{\cB}^{L m}) - (\cB^I \wt{\cB}^{L k})(\cB^{J k} \cB^{J m})\right), \\[5pt]
\dl \cM_3&=-\frac{2}{3}\, \ve^{IL} \D^m \left((\cB^J \cB^J)(\cB^I \wt{\cB}^{L m}) -2 (\cB^I \wt{\cB}^{L k})(\cB^{J k} \cB^{J m})\right), \\[5pt]
\dl \cM_4&=-4 \ve^{IL} \D^m (\cB^{I k} \cB^{J n})(\cB^{J n} \wt{\cB}^{L m k}), \\[5pt]
\dl \cM_5&=-2 \ve^{IL} \D^m (\cB^J \cB^J)(\cB^I \wt{\cB}^{L m}),\label{dm5} \\[5pt]
\dl \cM_6&=-4 \ve^{IL} \D^m (\cB^{Jkl} \cB^{Jnj})(\cB^{Ikn} \wt{\cB}^{Lmlj}). \label{dlA6}
\end{align}
The variations of $\cM_1$, $\cM_4$ and $\cM_6$ are straightforward.
The variation of $\cM_2$ follows from the identity
\be
\ve^{IJ}(\cB^{Jm} \wt{\cB}^{Iln})= \frac{1}{6}\,\ve^{IJ}\left((\cB^{I} \wt{\cB}^{Jn})\,\dl^{lm}-(\cB^{I} \wt{\cB}^{Jl})\,\dl^{nm}\right),
\ee
which can be derived by writing at its right hand side $\cB$ in terms of $\wt\cB$, and vice versa. To put $\dl\cM_5$ in the form \eref{dm5} it is sufficient to notice that $(\cB^I\wt\cB^{Jm})$ is antisymmetric in $I$ and $J$, so that $(\cB^I\wt\cB^{Jm})=\frac{1}{2}\,\ve^{IJ}\ve^{MN}(\cB^M\wt\cB^{Nm})$. The computation of
$\dl \cM_3$ is slightly more involved. Define the vector
\be
W^\m = \ve^{IJ} \ve^{\m \s \m_1 \m_2 \m_3 \n_1 \n_2 \n_3}  \cB^I_{\m_1 \m_2 \m_3} \cB^J_{\n_1 \n_2 \n_3}v_\s,
\ee
which for $v^\m=(1,0,\cdots,0)$ reduces to
\be
W^n=-6\ve^{IJ} \big(\cB^I \wt{\cB}^{J n}\big).
\ee
Contracting the two Levi-Civita tensors one finds that the square of $W^n$ reduces to a combination of the above polynomials, namely\footnote{Notice that, in $D=4$, the same calculation gives just the combination $W^\m W_\m=16Q_2= 2(\cM_5-\cM_1)$, see Footnote \ref{wd4} in Section \ref{canthpst}.}
\be\label{ww}
W^n W^n = -W^\m W_\m= 72(\cM_1-9\cM_2+9\cM_3-\cM_5).
\ee
The variation of $W^n$
\be
\dl W^n=12 \D^m (\wt{\cB}^{J m} \wt{\cB}^{J n})= 12 \D^m \left((\cB^J \cB^J)\dl^{m n}-3(\cB^{J m} \cB^{J n})\right),
\ee
then yields for the variation of its square
\be
\dl (W^n W^n)= -144 \,\ve^{IL} \D^m \left((\cB^J \cB^J)(\cB^I \wt{\cB}^{L m}) -3 (\cB^{J k} \cB^{J m})(\cB^I \wt{\cB}^{L k})\right).
\ee
Knowing the variations of $\cM_1$, $\cM_2$ and $\cM_5$, one can so determine  $\dl\cM_3$. As the variations of $\cM_4$ and $\cM_6$ contain terms that do not appear in the other variations, they cannot give rise to polynomials that satisfy the PST condition \eref{deltaa}. Conversely, with the other four polynomials we can form the two combinations
\be \label{s1s2}
S_1=\cM_5-\frac{1}{2}\,\cM_1,\quad\quad S_2=\cM_1-12\cM_2+6\cM_3,
\ee
which satisfy indeed $\dl S_1=0=\dl S_2$. The invariant $S_1$ corresponds to the canonical duality invariant quartic interaction \eref{canquarpst}, whereas $S_2$ is a new non-canonical one.

\subsection{Matching the PST and GZGR deformations}

Above we have determined the most general duality and Lorentz invariant quartic deformations of a non-linear Maxwell theory in both the GZGR and the PST formalisms, the polynomials $R_1$ and $R_2$ \eqref{inv2}, and $S_1$ and $S_2$ \eqref{s1s2}, respectively.
These results are new and interesting by themselves, but they also allow us to discuss the relation, and eventually the equivalence, of the two approaches in this specific case. The PST condition ensures that $S_1$ and $S_2$, once mapped to the GZGR side, become Lorentz invariant polynomials of $F$. This means that they necessarily go over into linear combinations of the polynomials  $\cK_1,\cdots,\cK_4$ in \eref{k1}-\eref{k4}. Conversely, what is not guaranteed is that duality invariant interactions of the PST approach, as are $S_1$ and $S_2$, go over to duality invariant interactions of the GZGR approach, namely to linear combinations of $R_1$ and $R_2$. In this section, we verify this
correspondence explicitly, thus supporting our conjecture -- in the literature sometimes implicitly assumed -- that the manifestly Lorentz invariant (GZGR) and the manifestly duality invariant (PST) formulations are eventually equivalent.

The check of this correspondence regards, actually, only $S_2$, as the polynomial $S_1$ is the canonical PST quartic interaction, that we know already to correspond to the canonical quartic GZGR interaction $R_1$, see equations \eref{canquar} and \eref{canquarpst}.
To map the polynomial $S_2$ to the GZGR side, we follow the general strategy of Method {\it II} outlined in Section \ref{gzgrpst}. In fact, according to the mapping \eref{karel}, it suffices to enforce in $S_1$ and $S_2$ the identifications
\be
\cB^1=\cE^2\equiv E, \quad\quad \cB^2\equiv B.
\ee
In this way, $S_1$ and $S_2$ go over into, see Appendix \ref{quartpg} for details,
\begin{align} \label{s1c}
S_1=&\,\frac{1}{2}\left((EE)^2+(BB)^2\right)+2(EB)^2-(EE)(BB),\\[5pt]
\begin{split} \label{s2c}
S_2=&\,(EE)^2+(BB)^2+2(EE)(BB)-6\left((E^iE^j)(E^iE^j)+(B^iB^j)(B^iB^j)\right) \\[2pt]
&-24(E^iE^j)(B^iB^j)+12(E^iB^j)(E^iB^j).
\end{split} \end{align}
Correspondingly, we must write out also $R_1$ and $R_2$ in terms of the fields $E$ and $B$, using \eref{ebff},
\begin{align}
R_1=&\,16\left((EE)^2+(BB)^2 +4(EB)^2 -2(EE)(BB)\right),\label{r1c}
\\[5pt]
\begin{split}
R_2=&\,2(EB)^2-(EE)^2-(BB)^2-4(EE)(BB)+36(E^i E^j)(B^i B^j)\\[2pt]
&-18(E^i B^j)(E^i B^j)+9\left((E^i E^j)(E^i E^j)+(B^i B^j)(B^i B^j)\right).
\end{split}\label{r2c}
\end{align}
In fact, since $S_1$ and $R_1$ are proportional to the canonical invariants, formulas \eref{s1c} and \eref{r1c} can be read off directly from \eref{canquarpst} and \eref{canquar}, respectively. From the above expressions one sees that $S_1$ and $S_2$ go indeed over to combinations of only the duality invariant polynomials $R_1$ and $R_2$,
\be\label{sr12}
S_1=\frac{1}{32}\,R_1, \quad\quad S_2=-\frac{2}{3}\,R_2+\frac{1}{48}\,R_1.
\ee
This implies that the generic quartic deformation $\cM=aS_1+bS_2$ of the PST approach corresponds, in the GZGR approach, to the deformation (apply \eref{karel} with $n=4$)
\[
\cK=\frac{8}{3}\,bR_2-\frac{1}{24}\,(3a+2b)R_1.
\]

\section{Conclusions and outlook}\label{out}

The GZGR approach for duality invariant self-interactions of Maxwell fields in higher dimensional space-times features, as principal advantage, manifest Lorentz invariance, realized in a standard way. Conversely, $SO(2)$-duality and $S$-duality invariance are realized, at the level of equations of motion, by means of a non-linear differential equation on the Lagrangian $\cL(F)$, the GZGR condition \eref{gzgr1}. The PST approach has the advantage of realizing both dualities in a manifest way. In particular, $SO(2)$ is realized as a Noether symmetry of the action. In turn, the implementation of Lorentz invariance -- alias the condition of non-propagation of the auxiliary field $a$ -- imposes restrictions on the Hamiltonian $\cN(\cB)$: as we have shown in this paper, the general condition the Hamiltonian must satisfy is the PST differential equation \eref{pst22}.

Although there is a one-to-one correspondence between the (not  necessarily Lorentz and/or duality invariant) functions $\cL(F)$ and $\cN(\cB)$, there is no simple direct relation between the properties of the two formulations. We gave a proof that if $\cN(\cB)$ satisfies the PST condition, then $\cL(F)$ is Lorentz invariant, and vice versa. What is in general not guaranteed is that the GZGR condition implies duality invariance on the PST side, and vice versa. However, if we restrict the functional dependencies of $\cL(F)$ and $\cN(\cB)$ to canonical quadratic and quartic invariants, as we have shown, the two approaches are completely equivalent: although the resulting consistency conditions \eref{ch1} and \eref{ch3} are formally the same, being in turn equivalent to the Courant-Hilbert equation, the proof is highly non-trivial. In particular, the algebraic forms of the conditions for Lorentz and duality invariance in the two formulations are interchanged, making the relation non-intuitive.

A comparison between the two approaches going beyond canonical interactions has to cope with the technical problem that the respective consistency conditions \eref{gzgr1} and \eref{pst22} can hardly be solved in a closed form, and look formally rather different. The former is a {\it scalar} equation, ensuring invariance under the one-parameter group $SO(2)$, whereas the latter amounts to an equation for a spatial {\it vector}, see for instance \eref{deltaa}, ensuring essentially invariance under {\it special} Lorentz transformations in $D$ dimensions. In light of these general difficulties, we investigated the equivalence problem for the simplest, even though technically rather involved, non-canonical interactions, namely the most general duality invariant quartic interactions for a  Maxwell theory in $D=8$. It turned out that, on the GZGR side, there is one non-canonical (manifestly Lorentz invariant) duality invariant quartic interaction, while on the PST side there is one non-canonical  (manifestly duality invariant) Lorentz invariant such interaction: we found that the two interactions describe the same physical dynamics. Despite the poor ``statistical" significance of this result, we take this highly non-trivial check for a strong indication that the GZGR and PST formulations are, eventually, fully equivalent.

To test the GZGR/PST equivalence on a deeper level, one would need a general algorithm to solve the consistency conditions \eref{gzgr1} and \eref{pst22} order by order. Taking into account the next order $F^6$, the Hamiltonian would have the form $\cN(\cB)=\frac{1}{2}\cB^I \cB^I + \cM_{(4)}(\cB) + \cM_{(6)}(\cB)$, where the quartic polynomial is given by $\cM_{(4)}= aS_1+bS_2$, see equations \eref{sr12}, and satisfies $\dl\cM_{(4)}=0$. The PST condition \eref{pstalin} then translates into the equation for the sixth-order polynomial
\be\label{six}
\dl\cM_{(6)}=\frac{1}{2\cdot3!}\,\Delta_\s\,\ve^{IJ}\ve^{\rho\s\m_1\m_2\m_3\n_1\n_2\n_3} M^I_{(4)\m_1\m_2\m_3}\,M^J_{(4)\n_1\n_2\n_3}v_\rho.
\ee
As in the case of the canonical invariants
\eref{nq12}, this equation must allow for a ``particular'' solution $\w\cM_{(6)}$, determined by the form of the quartic polynomial $\cM_{(4)}$, but then, in principle, one must add the most general solution of the homogeneous equation $\dl\cM_{(6)}=0$. This homogeneous equation, for sixth order polynomials, has no solutions for canonical invariants, as one sees from the universal solution \eref{nq12}. However, we cannot exclude their existence for non-canonical sixth-order invariants. An analogous analysis applies for the GZGR condition \eref{gzgr1}. As seen in the text, the algebra involved for the analysis of the quartic polynomials is already rather complex, and to determine the most general solution of the equation $\dl\cM_{(4)}=0$ we had to resort to some ingenious algebraic identities. To test the equivalence hypothesis in the more complex situation of sixth-order polynomials, one would need a more systematic algebraic algorithm, still to be developed.

The inclusion of derivative couplings in the present framework presents no conceptual difficulties. In the GZGR formulation, one must replace equation \eref{gzgr1} with the integral form $\int(F\wt F-L\wt L)\,d^Dx=0$, where now $F$ is no longer considered as the field strength of a potential, but as an independent field, and $L_{\m_1\cdots\m_n}
=(\dl/\dl F^{\m_1\cdots\m_n})\,\int \cL(F)\,d^Dx$, see e.g. \cite{AF}. On the PST side, the consistency condition \eqref{pst22} remains formally the same, apart from the fact that the tensors $N^I$ are now functional derivatives,  $N^I_{\m_1\cdots\m_p}
=(\dl/\dl \cB^{I\m_1\cdots\m_n})\,\int \cN(\cB)\,d^Dx$.
This distinctive feature between the two formulations is not unexpected, as the two conditions ensure duality invariance and Lorentz invariance, respectively.
Explicit duality invariant examples of derivative theories are, however, rather rare, even in $D=4$; for an example of a quadratic theory, see, for instance, \cite{BN,AF}. The explicit construction of duality invariant non-linear derivative theories in dimensions $D\ge 8$ is at the moment a completely unexplored field.

When non-linear duality invariant Maxwell theories are coupled to charged sources, there is a major difference arising between the two formulations, which is, actually, inherited from the linear theory. In the GZGR formulation, although the equations of motion are invariant under $SO(2)$ rotations, the quantization condition for the charges -- Dirac's original condition \eref{Dirac} -- is not invariant under $SO(2)$. In fact, the related action does not possess an $SO(2)$ symmetry, not even at the formal level, i.e. if one rotates also the charges. In turn, this formulation entails the phenomenon of spin-statistics transmutation. Conversely, the quantization condition arising in the PST formulation -- Schwinger's condition
\eref{Schwinger} -- is $SO(2)$-invariant, as is the related action, and, correspondingly, it does not entail spin-statistics transmutation. In a truly $SO(2)$ invariant theory there is, in fact, no elementary self-interaction of the $r$-th dyonic $(n-2)$-brane with itself.

Four-dimensional duality invariant Maxwell theories are intimately related with the dynamics of chiral two-form potentials in a six-dimensional space-time. A double dimensional reduction of the latter theory produces a manifestly duality invariant Maxwell theory in $D=4$, formulated \`a la PST, as the Lorentz group splits into $SO(1,5)\ra SO(1,3)\times SO(2)$, see \cite{Berman,Nurma} for the case of the Born-Infeld theory. A similar relation is expected to hold in general between chiral $2N$-form potentials in  $D=4N+2$, and duality invariant Maxwell theories in $D=4N$. The simplest higher-dimensional case corresponds to chiral four-forms in $D=10$, whose non-linear dynamics should thus give rise to the non-linear (canonical and/or non-canonical) duality invariant Maxwell theories in $D=8$ constructed in this paper. There is, however, no natural extension of a Born-Infeld-like dynamics or, more generally, of a ``canonical'' dynamics, for a chiral four-form in $D=10$. The analysis of this interesting issue will be presented elsewhere \cite{BLM}.

\vskip0.5truecm
\paragraph{Acknowledgements.}
This work is supported in part by the INFN CSN4 Special Initiative {\it STEFI}. K.L. is grateful to Dmitri Sorokin for a series of enlightening discussions.

\appendix
\addcontentsline{toc}{section}{APPENDICES}

\section{Linking the PST and GZGR formulations of canonical theories}\label{link}

Establishing the link between the two approaches requires to establish an explicit relation between the derivatives $\cN_i$ and $\cL_i$\footnote{This appendix presents also some details of the calculations not given in reference \cite{DS}, on which it is mainly based.}. As first step, we write out the canonical variables on the PST side
\be\label{xyz}
Q_1=\frac{1}{2}\,\cB^I\cB^I,\quad\quad Q_2=
\frac{1}{4}\left((\cB^1\cB^2)^2-(\cB^1\cB^1)(\cB^2\cB^2)\right),\quad\quad Q_3=\frac{1}{2}\,\cB^2\cB^2,
\ee
and on the GZGR side
\be\label{alfabeta}
P_1=\frac{n}{2}\left(\cE^2 \cE^2-\cB^2\cB^2\right), \quad\quad P_2=\frac{n^2}{4}\left(\cE^2\cB^2\right)^2,\quad\quad P_3=-\frac{n}{2}\,\cB^2\cB^2,
\ee
where we enforced the identification $F=F^2$, leading to the equality $P_3=-nQ_3$.
To connect the derivatives $\cN_i$ to $\cL_i$, we use the constitutive relation between $\cN$ and $\cL$ \eref{lagf2}
\be
n\cN=n\cE^2\cB^1-\cL,
\ee
and write out the conjugate variable of $\cE^2$ (see \eref{simplec})
\be\label{b1e2}
\cB^1= \frac{1}{n}\,\frac{\pa\cL}{\pa\cE^2}=\cL_1\,\cE^2+\frac{n}{2}\,\cL_2\left(\cE^2\cB^2\right)
\cB^2.
\ee
Using this expression, we can write out the variables $Q_1$ and $Q_2$, and the Hamiltonian
\begin{align}
\label{defx}Q_1&=\left(1+\cL_1^2+P_2\cL_2^2\right)Q_3+\frac{1}{n}\, P_1\cL_1^2+\frac{2}{n}\,P_2\cL_1\cL_2,\\[5pt]
\label{defy}Q_2&=\left(\frac{1}{n^2}\,P_2-\frac{1}{n}\,Q_3P_1
-Q_3^2\right)\cL_1^2,\\[5pt]
\label{defn}n\cN&=2(P_1+nQ_3)\cL_1+2P_2\cL_2-\cL.
\end{align}
Computing the differential of the Hamiltonian we find
\be\label{dn}
nd\cN=\cL_1dP_1+\cL_2dP_2+2(P_1+nQ_3)\,d\cL_1+2P_2 d\cL_2+n\left(2\cL_1+\cL_3\right) dQ_3.
\ee
Similarly, it is straightforward to express the differentials $dQ_1$ and $dQ_2$ in terms of $dP_1$, $dP_2$, $d\cL_1$, $d\cL_2$, $dQ_3$, and furthermore to compute the combination
\begin{align*}
dQ_1-n\,\frac{\cL_2}{\cL_1}\,dQ_2 =&\left(Q_3\cL_2+ \frac{1}{n}\, \cL_1\right)\big(\cL_1dP_1+\cL_2dP_2 +2(P_1+nQ_3)\,d\cL_1+2P_2 d\cL_2\big) +\\[5pt]
&\left(1+\cL_1^2+P_2\cL_2^2+(P_1+2nQ_3)\,\cL_1\cL_2\right)dQ_3.
\end{align*}
Comparing this relation with \eref{dn}, we eventually find an explicit expression for the differential of $\cN(Q)$
\be
d\cN=\frac{1}{\cL_1+nQ_3\cL_2}\,\left(dQ_1-n\,\frac{\cL_2}{\cL_1}\,dQ_2+\left(\cL_1^2-P_1\cL_1\cL_2-P_2\cL_2^2
+\left(\cL_1+nQ_3\cL_2\right)\cL_3-1\right)dQ_3
\right).
\ee
Thus, we recover the relations between the derivatives we were looking for
\be\label{nxyzab}
\begin{split}
\cN_1&=\frac{1}{\cL_1+nQ_3\cL_2},\\[5pt]
\cN_2&=-\frac{n\cL_2/\cL_1}{\cL_1+nQ_3\cL_2},\\[5pt]
\cN_3&=\frac{\cL_1^2-P_1\cL_1\cL_2-P_2\cL_2^2+\left(\cL_1+nQ_3\cL_2\right)\cL_3-1}{\cL_1+nQ_3\cL_2}.
\end{split}
\ee
Using these formulas, and inserting for $Q_1$ and $Q_2$ the expressions \eref{defx} and \eref{defy}, we find the identity
\be\label{id123}
\cN_1^2 -Q_1\cN_1\cN_2-Q_2\cN_2^2+\left(\cN_1-Q_3\cN_2\right)\cN_3-1=\frac{\cL_3}{\cL_1},
\ee
that parallels the third relation in \eref{nxyzab}. From formulas \eref{nxyzab} we also derive that
\be
\cN_1-Q_3\cN_2=\frac{1}{\cL_1}.
\ee
Using this relation, and the above expression of $\cN_3$, \eref{id123} implies the further identity
\be\label{equiv}
\frac{\cL_1^2-P_1\cL_1\cL_2- P_2\cL_2^2-1}{\cL_1}=-\frac
{\cN_1^2-Q_1\cN_1\cN_2- Q_2\cN_2^2-1}{\cN_1}.
\ee

\section{Proof of a tensor identity}\label{idencom}

We did not find any direct way to prove the identity \eref{compl}. An indirect proof can be constructed as follows. Define the two quartic invariants ($X$ is just the invariant \eref{compl} we are interested in)
\be
X=\big(F^{\m\n} F^{\rho\s}\big)\big(F_{\m\n} \wt F_{\rho\s}\big), \quad\quad Y=
\big(F^{\m\n} F^{\rho\s}\big)\big(F_{\m\rho} \wt F_{\n\s}\big).
\ee
Consider $Y$ and write the second factor $F^{\rho\s\g_1\g_2}$ as minus the Hodge dual of $\wt F_{\a_1\a_2\a_3\a_4}$, and express the last factor $\wt F_{\n\s\dl_1\dl_2}$ in terms of $F_{\bt_1\bt_2\bt_3\bt_4}$,
\be\label{ap1}
Y=-\frac{1}{(4!)^2}\,F^{\m\n}{}_{\g_1\g_2}\,\ve^{\rho\s\g_1\g_2\a_1\a_2\a_3\a_4} \,\wt F_{\a_1\a_2\a_3\a_4}\,F_{\m\rho}{}^{\dl_1\dl_2}\,\ve_{\n\s\dl_1\dl_2\bt_1\bt_2\bt_3\bt_4} F^{\bt_1\bt_2\bt_3\bt_4}.
\ee
As the Levi-Civita tensors have one index in common, and the first and third $F$ have the index $\m$ contracted, there remains a contraction of seven completely antisymmetrized indices, which can be written as
\be
Y=\frac{7!}{(4!)^2}\,\left(F^{[\m_1}{}_{\m_1\m_2}\,F^{\m_2\m_3}{}_{\m_3}\right)
F^{\m_4\m_5\m_6\m_7]}\,\wt F_{\m_4\m_5\m_6\m_7}.
\ee
It is a now a mere, though a bit lengthy, exercise to write out the antisymmetrization and to contract the indices. The result reproduces a combination of $Y$, $X$ and $(FF)(F \wt F)$, namely
\be
Y=Y+\frac{1}{2}\,X-\frac{1}{24}\,(FF)(F\wt F),
\ee
which amounts to the identity \eref{compl}. Alternatively, one may apply a similar procedure starting from the invariant $X$, by writing the first factor $F^{\m\n\g_1\g_2}$ as minus the
Hodge dual of $\wt F_{\a_1\a_2\a_3\a_4}$. In this case the computation is even more lengthy, as one remains with the antisymmetrization of eight indices. This time $Y$ cancels out, and the result is
\[
X=2X-\frac{1}{12}\,(FF)(F\wt F),
\]
which corresponds again to \eref{compl}.

On the other hand, the invariant $Y$ cannot be proportional to the polynomial $(FF)(F\wt F)$. In fact, would it be so, then also the combination $X-2Y$ would be proportional to this polynomial
\be\label{xyff}
X-2Y=3\big(F^{[\m\n} F^{\rho\s]}\big)\big(F_{\m\n} \wt F_{\rho\s}\big)=c (FF)(F\wt F).
\ee
Consider now a field configuration for which the only non-vanishing components are $\a=F^{0123}$ and $\bt =F^{4567}$. Then the right hand side of \eref{xyff} is non-vanishing, actually proportional to $c\a\bt(\a^2-\bt^2)$, while the left hand side is zero. It follows that $Y$ cannot be proportional to $(FF)(F\wt F)$ (unless $X-2Y$ is identically zero, but from the first equality in \eref{xyff} it is easily seen that this is not the case).

\section{Quartic polynomials in the PST and GZGR formulations}\label{quartpg}

Setting in the relevant invariants in \eqref{a1}-\eqref{a6} $\cB^1= E$ and $\cB^2= B$, one obtains
\begin{align}
\cM_1&=(EE)^2+(BB)^2+2(EE)(BB),\label{a11} \\[5pt]
\cM_2&=(E^iE^j)(E^iE^j)+(B^iB^j)(B^iB^j)+2(E^iE^j)(B^iB^j), \\[5pt]
\cM_3&=(E^iE^j)(E^iE^j)+(B^iB^j)(B^iB^j)+2(E^iB^j)(E^iB^j), \\[5pt]
\cM_5&=(EE)^2+(BB)^2+2(EB)^2 \label{a55}.
\end{align}
Inserting these expressions in the two independent PST-invariant combinations in \eqref{s1s2} then yields \eref{s1c}, and \eref{s2c}. To write out the GZGR invariants in \eref{inv2} in terms of the electric and magnetic fields, we specify the definitions \eref{ebff} to $D=8$
\be
F^{i_1\cdots i_30}=E^{i_1\cdots i_3}, \quad\quad F^{i_1\cdots i_4}=-\frac{1}{3!}\ve^{i_1\cdots i_7}B^{i_5\cdots i_7}.
\ee
The results for $\cK_1$ and $(F\wt F)^2$ are trivial
\be
\cK_1=16(EE-BB)^2, \quad\quad (F\wt F)^2=64(EB)^2.
\ee
For $\cK_2$, the computation is slightly more involved and leads to
\be
\cK_2=(F^0 F^0)(F_0 F_0)+2(F^0 F^i)(F_0 F_i)+(F^i F^j)(F_i F_j),
\ee
where
\begin{align}
(F^0 F^0)(F_0 F_0)=&\,(EE)^2,\\[5pt]
(F^0 F^i)(F_0 F_i)=&\,(EB)^2-(EE)(BB)+9\left((E^i E^j)(B^i B^j)-(E^i B^j)(E^i B^j)\right),\\[5pt]
\begin{split}
(F^i F^j)(F_i F_j)=&\,9\left((E^i E^j)(E^i E^j)+(B^i B^j)(B^i B^j)+2(E^i E^j)(B^i B^j)\right)\\
&+(BB)^2-6(EE)(BB).
\end{split}
\end{align}
Thus, we obtain
\be \begin{split}
\cK_2=&\,(EE)^2+(BB)^2-8(EE)(BB)+2(EB)^2+36(E^i E^j)(B^i B^j)\\
&-18(E^i B^j)(E^i B^j)+9\left((E^i E^j)(E^i E^j)+(B^i B^j)(B^i B^j)\right).
\end{split} \ee
Finally, the two GZGR invariant combinations in \eref{inv2} become
\begin{align}
R_1=&\,16\left((EE)^2+(BB)^2-2(EE)(BB)+4(EB)^2\right),
\\[10pt]
\begin{split}
R_2=&\,2(EB)^2-(EE)^2-(BB)^2-4(EE)(BB)+36(E^i E^j)(B^i B^j)\\[2pt]
&-18(E^i B^j)(E^i B^j)+9\left((E^i E^j)(E^i E^j)+(B^i B^j)(B^i B^j)\right),
\end{split}
\end{align}
which are formulas \eref{r1c}, \eref{r2c}.

%

\vskip0.5truecm
\begin{thebibliography}{99}


\bibitem{DGNW} S. Deser, M.T. Grisaru, P. van Nieuwenhuizen and C.C. Wu, {\it Scale dependence and the renormalization problem of quantum gravity}, Phys. Lett. {\bf B58} (1975) 355.

\bibitem {Nov} J. Novotny, {\it Self-duality, helicity conservation and normal ordering in nonlinear QED}, Phys. Rev. {\bf D98} (2018) 085015,
arXiv:1806.02167 [hep-th].

\bibitem{DT} S. Deser and C. Teitelboim, {\it Duality transformations of abelian and non-abelian gauge fields}, Phys. Rev. {\bf D13} (1976) 1592.

\bibitem{PST1} P. Pasti, D. Sorokin and M. Tonin, {\it Covariant actions for models with non-linear twisted self-duality}, Phys. Rev. {\bf D86} (2012) 045013, arXiv:1205.4243 [hep-th].

\bibitem{BN} G. Bossard and H. Nicolai, {\it Counterterms vs. dualities},  JHEP {\bf 1108} (2011) 074, arXiv:1105.1273 [hep-th].

\bibitem{EHJP}  H. Elvang, M. Hadjiantonis, C.R.T. Jones and S. Paranjape, {\it All-multiplicity one-loop amplitudes in Born-Infeld electrodynamics from generalized unitarity}, arXiv:1906.05321 [hep-th].

\bibitem{GZ1} M.K. Gaillard and B. Zumino, {\it Duality rotations for interacting fields}, Nucl.
     Phys. {\bf B193} (1981) 221.

\bibitem{GZ2} M.K. Gaillard and B. Zumino, {\it Non-linear electromagnetic self-duality and Legendre transformations}, In: Duality and Supersymmetric Theories, eds. D.I. Olive and P.C. West, p.33, Cambridge University Press (1999), hep-th/9712103.

\bibitem{GR} G.W. Gibbons and D.A. Rasheed, {\it Electric-magnetic duality rotations in non-linear electrodynamics}, Phys. Lett. {\bf B365} (1996) 46.

\bibitem{IZ1} E.A. Ivanov and B.M. Zupnik, {\it New representation for Lagrangians of self-dual nonlinear electrodynamics}, In: Supersymmetries and quantum symmetries, eds. E.A. Ivanov et al, p. 235, Dubna (2002), hep-th/0202203.

\bibitem{IZ2} E.A. Ivanov and B.M. Zupnik, {\it New approach to nonlinear electrodynamics: Dualities as symmetries of interaction}, Yadern Fiz. {\bf 67} (2004) 2212; [Phys. Atom. Nucl. {\bf 67} (2004) 2188], hep-th/0303192.

\bibitem{IZ3} E.A. Ivanov and B.M. Zupnik, {\it Bispinor auxiliary fields in duality-invariant electrodynamics revisited}, Phys. Rev. {\bf D87} (2013) 065023, arXiv:1212.6637 [hep-th].

\bibitem{Kuz1} S.M. Kuzenko, {\it Manifestly duality-invariant interactions in diverse dimensions}, arXiv:1908.04120 [hep-th].

\bibitem{KuTh1} S.M. Kuzenko and S. Theisen, {\it Supersymmetric duality rotations}, JHEP {\bf 0003} (2000) 034, hep-th/0001068.

\bibitem{KuTh2} S.M. Kuzenko and S. Theisen, {\it Nonlinear self-duality and supersymmetry}, Fortsch. Phys. {\bf 49} (2001) 273,  hep-th/0007231.

\bibitem{HT} M. Henneaux and C. Teitelboim, {\it Dynamics of chiral (self-dual) p-forms}, Phys. Lett.  {\bf B206} (1988) 650.

\bibitem{SS} J.H. Schwarz and A. Sen, {\it Duality symmetric actions},  Nucl. Phys. {\bf B411} (1994) 35, hep-th/9304154.
 and \"O. Sarioglu, {\it Hamiltonian electric/magnetic duality and Lorentz invariance}, Phys.

\bibitem{DGHT1} S. Deser, A. Gomberoff, M. Henneaux and C. Teitelboim, {\it Duality, self-duality, sources and charge quantization in abelian N-form theories}, Phys. Lett. {\bf B400} (1997) 80, hep-th/9702184.

\bibitem{DGHT2} S. Deser, A. Gomberoff, M. Henneaux and C. Teitelboim, {\it $p$-brane dyons and electric-magnetic duality},  Nucl. Phys. {\bf B520} (1998) 179,  hep-th/9712189.

\bibitem{DS} S. Deser and \"O. Sarioglu, {\it Hamiltonian electric/magnetic duality and Lorentz invariance}, Phys. Lett. {\bf B423} (1998) 369, hep-th/9712067.

\bibitem{BC} X. Bekaert and S. Cucu, {\it Deformations of duality-symmetric theories}, Nucl. Phys. {\bf B610} (2001) 433, hep-th/0104048.

\bibitem{BH} X. Bekaert and M. Marc Henneaux, {\it Comments on chiral p-forms}, Int. J. Theor. Phys. {\bf 38} (1999) 1161, hep-th/9806062.

\bibitem{PamD} P.A.M. Dirac, {\it The Conditions for a quantum field theory to be relativistic}, Rev. Mod. Phys. {\bf 34} (1962) 592.

\bibitem{Schw} J. Schwinger, {\it Commutation relations and conservation laws}, Phys. Rev. {\bf 130} (1963) 406.

\bibitem{PST2} P. Pasti, D. Sorokin and M. Tonin, {\it Duality symmetric actions with manifest
space-time symmetries}, Phys. Rev. {\bf D52} (1995) 4277, hep-th/9506109.

\bibitem{PST22} P. Pasti, D. Sorokin and M. Tonin, {\it Space-time symmetries in duality symmetric models}, in {\it Gauge Theories, applied supersymmetry and quantum gravity}, proceedings, Leuven, Belgium (1995), hep-th/9509052.

\bibitem{PST3} P. Pasti, D. Sorokin and M. Tonin, {\it On Lorentz invariant actions for chiral $p$-forms}, Phys. Rev. {\bf D55} (1997) 6292, hep-th/9611100.

\bibitem{PST4} P. Pasti, D. Sorokin and M. Tonin, {\it Covariant action for a $D=11$ 5brane with the chiral field}, Phys. Lett. {\bf B398} (1997) 41, hep-th/9701037.

\bibitem{DLS} G. Dall'Agata, K. Lechner and D. Sorokin, {\it Covariant actions for the bosonic sector of $d = 10$ IIB supergravity}, Class. Quant. Grav. {\bf 14} (1997) L195, hep-th/9707044.

\bibitem{BLNPST} I.A. Bandos, K. Lechner, A. Nurmagamebetov, P. Pasti, D. Sorokin and M. Tonin, {\it Covariant action for the superfive-brane of $M$ theory}, Phys. Rev. Lett. {\bf 78} (1997), hep-th/9701149.

\bibitem{DLT1} G. Dall'Agata, K. Lechner and M. Tonin, {\it Covariant actions for $N=1$,
      $D = 6$ supergravity theories with chiral bosons}, Nucl. Phys. {\bf B512} (1998) 179,  hep-th/9710127.

\bibitem{DLT2} G. Dall'Agata, K. Lechner and M. Tonin, {\it $D = 10$, $N = IIB$ supergravity: Lorentz invariant actions and duality}, JHEP {\bf 9807} (1998) 017,  hep-th/9806140.

\bibitem{L} K. Lechner, {\it Self-dual tensors and gravitational anomalies in 4n + 2-dimensions}, Nucl. Phys. {\bf B537} (1999) 361, hep-th/9808025.

\bibitem{LM} K. Lechner and P.A. Marchetti, {\it Duality invariant quantum field theories of charges and monopoles}, Nucl. Phys. {\bf B569} (2000) 529, hep-th/9906079.

\bibitem{CKR} J.J.M. Carrasco, R. Kallosh and R. Roiban, {\it Covariant procedures for perturbative non-linear deformation of duality-invariant theories}, Phys. Rev. {\bf D85} (2012) 025007, arXiv:1108.4390 [hep-th].

\bibitem{CH} R. Courant and D. Hilbert, {\it Methods of Mathematical Physics}, Vol.
II, Interscience (1962) 91.

\bibitem{BI} M. Born and L. Infeld, {\it Foundations of the new field theory}, Proc. Roy. Soc. {\bf A144} (1934) 425.

\bibitem{Born} M. Born, {\it Th\'eorie non-lin\'eaire du champ \'electromagn\'etique}, Ann. Inst. Poincar\'e {\bf 7} (1937) 155.

\bibitem{Tan} Y. Tanii, {\it Introduction to supergravities in diverse dimensions}, hep-th/9802138.

\bibitem{ArTan} M. Araki and Y. Tanii, {\it  Duality symmetries in non-linear gauge theories},
Int. J. Mod. Phys. {\bf A14} (1999) 1139, hep-th/9808029.

\bibitem{ABMZ} P. Aschieri, D. Brace, B. Morariu and B. Zumino, {\it Nonlinear self-duality in even dimensions}, Nucl. Phys. {\bf B574} (2000) 551, hep-th/9909021.

\bibitem{Berman} D. Berman, {\it SL(2,Z) duality of Born-Infeld theory from non-linear
self-dual electrodynamics in six dimensions}, Phys. Lett. {\bf B409} (1997) 153, hep-th/9706208.

\bibitem{Nurma} A. Nurmagambetov, {\it Duality symmetric three-brane and its coupling to type IIB supergravity}, Phys. Lett. {\bf B436} (1998) 289, hep-th/9804157.

\bibitem{BLM} G. Buratti, K. Lechner and L. Melotti, in preparation.

\bibitem{AFT} P. Aschieri, S. Ferrara and S. Theisen, {\it Constitutive relations, off shell duality rotations and the hypergeometric form of Born-Infeld theory}, Springer Proc. Phys. {\bf 153} (2014) 23, arXiv:1310.2803 [hep-th].

\bibitem{lege} A.N. Iusem, D. Reem and S. Reich, {\it Fixed points of
Legendre-Fenchel type transforms}, J. of Convex Analysis {\bf 26} (2019) 275, arXiv:1708.00464 [math.CA].

\bibitem{PS} M. Perry and J.H. Schwarz, {\it Interacting chiral gauge fields in six dimensions and Born-Infeld theory}, Nucl. Phys.  {\bf B489} (1997) 47, hep-th/9611065.

 \bibitem{HKS} M. Hatsuda, K. Kamimura and S. Sekiya, {\it Electric magnetic duality invariant Lagrangians}, Nucl. Phys. {\bf B561} (1999) 341, hep-th/9906103.

\bibitem{LM1} K. Lechner and P.A. Marchetti, {\it Interacting branes, dual branes, and dyonic branes: a unifying Lagrangian approach in D dimensions}, JHEP {\bf 0101} (2001) 003, hep-th/0007076.

\bibitem{LM2} K. Lechner and P.A. Marchetti, {\it Spin-statistics transmutation in relativistic quantum field theories of dyons},  JHEP {\bf 0012} (2000) 028,  hep-th/0010291.

\bibitem{AF} P. Aschieri and A. Ferrara, {\it Constitutive relations and Schr\"odinger's formulation of nonlinear electromagnetic theories}, JHEP {\bf 1305} (2013) 087, arXiv:1302.4737 [hep-th].

\bibitem{ARN} I. Agullo, A. del Rio and J. Navarro-Salas, {\it Classical and quantum aspects of electric-magnetic duality rotations in curved spacetimes}, Phys. Rev. {\bf D98} (2018) 125001, arXiv:1810.08085 [gr-qc].


\end{thebibliography}
%

\end{document}